\documentclass[aps, prd, reprint, groupedaddress, amsmath, amssymb]{revtex4-2}
\usepackage{graphicx}
\usepackage{mathrsfs}
\usepackage{bm}
\usepackage{amssymb}
\usepackage{amsmath}
\usepackage{epsfig}

\usepackage{color}
\usepackage{mathtools}
\usepackage{cases}
\usepackage{booktabs}
\usepackage{natbib}

\usepackage{caption}
\usepackage{subcaption}
\usepackage{float}

\usepackage{tabularx}

\usepackage[
colorlinks=true, 
filecolor=black, 
anchorcolor=blue, 
linkcolor=blue, 
citecolor=cyan, 
urlcolor=cyan, 
linktocpage=true, 
plainpages=false, 
breaklinks=true, 
pdfstartview=FitH
]{hyperref}

\def\Mp{M_{P}}
\def\fh{f_{T}}
\def\fx{f_{X}}

\def\dd{\mathrm{d}}

\def\veck{\vec{k}}

\begin{document}

\title{Detectability of the chiral gravitational wave background \\ from audible axions with the LISA-Taiji network}

\author{Hong Su$^{1,5,6}$}
\author{Baoyu Xu$^{2,7}$}

\author{Ju Chen$^{3,5}$}
\email
{chenju@ucas.ac.cn}
\author{Chang Liu$^{4}$}
\email
{liuchang@yzu.edu.cn}
\author{Yun-Long Zhang$^{2,1}$}
\email
{zhangyunlong@nao.cas.cn}

\affiliation{$^{1}$ School of Fundamental Physics and Mathematical Sciences,  Hangzhou   Institute for Advanced Study, UCAS, Hangzhou 310024, China.}
\affiliation{$^{2}$National Astronomical Observatories, Chinese Academy of Sciences,
Beijing 100101, China}
\affiliation{$^{3}$ International Center for Theoretical Physics Asia-Pacific (ICTP-AP), University of Chinese Academy of Sciences, Beijing 100190, China}
\affiliation{$^{4}$Center for Gravitation and Cosmology, College of Physical Science and Technology, Yangzhou University, Yangzhou, 225009, China}
\affiliation{$^{5}$Taiji Laboratory for Gravitational Wave Universe (Beijing/Hangzhou), University of Chinese Academy of Sciences, Beijing 100049, China}
\affiliation{$^{6}$CAS Key Laboratory of Theoretical Physics, Institute of Theoretical Physics, Chinese Academy of Sciences, Beijing 100190, China.} 
\affiliation{$^{7}$
School of Astronomy and Space Science, University of Chinese Academy of Sciences, Beijing 100049, China.}

\date{\today}


\begin{abstract}
The chiral gravitational wave background (GWB) can be produced by axion-like fields in the early universe. We perform parameter estimation for two types of chiral GWB with the LISA-Taiji network: axion-dark photon coupling and axion-Nieh-Yan coupling. We estimate the spectral parameters of these two mechanisms induced by axion and determine the normalized model parameters using the Fisher information matrix. For highly chiral GWB signals that we choose to analyze in the mHz band, the normalized model parameters are constrained with a relative error less than $6.7\%$ (dark photon coupling) and $2.2\%$ (Nieh-Yan coupling) at the one-sigma confidence level. The circular polarization parameters are constrained with a relative error around $21\%$ (dark photon coupling) and $6.2\%$ (Nieh-Yan coupling) at the one-sigma confidence level.
 
\end{abstract}

\maketitle
\tableofcontents
\allowdisplaybreaks

\section{Introduction}
\label{section:1}  


The direct detection~\cite{LIGOScientific:2016aoc} of gravitational waves (GWs) by the Laser Interferometer Gravitational-Wave Observatory (LIGO)~\cite{Abramovici:1992ah} has offered a novel method for exploring the physics of the early Universe~\cite{Allen:1996vm,Maggiore:1999vm,Kuroyanagi:2018csn,LISACosmologyWorkingGroup:2022jok}. GWs produced by axions or axion-like particles (ALPs), especially the stochastic gravitational wave background (SGWB) from the early Universe, enable the detection of new physics beyond the Standard Model and provide insights into the early Universe~\cite{Binetruy:2012ze,Thrane:2013oya,Romano:2016dpx,Caprini:2018mtu,Boileau:2020rpg,Flauger:2020qyi,vanRemortel:2022fkb}. Axions were originally introduced to address the strong CP problem within the Standard Model~\cite{Peccei1977hh,Peccei:1977ur,Weinberg:1977ma,Wilczek:1977pj,ADMX:2018gho,DiLuzio:2020wdo}.  While numerous mechanisms exist for the production of axions in the early Universe~\cite{Abbott:1982af,Ipser:1983mw}, enabling a wide range of dark matter axion masses, these mechanisms may also contribute to various cosmological phenomena~\cite{Marsh:2015xka}. 

Axions and ALPs typically have weak couplings to photons or other Standard Model particles, making them difficult to detect directly~\cite{Preskill:1982cy,Sikivie:1983ip}. Moreover, these particles have been proposed to address other Standard Model and cosmological challenges, such as resolving the electroweak hierarchy problem~\cite{Georgi:1974yf,Graham2015cka}, serving as dark matter (DM) candidates~\cite{Preskill:1982cy,dine1983not,Bertone:2016nfn} or inflatons~\cite{Freese1990rb}, and being present in string theory frameworks~\cite{Arvanitaki:2009fg}. The audible axions model proposed in Refs.~\cite{Machado:2018nqk,Madge:2021abk} describes the coupling between axions and dark photons (gauge bosons), in which dark photons experience tachyonic instability when axions oscillate. The model postulates that axions or ALPs possess large initial velocities, enabling the generation of detectable GW signals even with small decay constants. This process results in the generation of an SGWB in the early Universe, allowing us to detect these particles, which carry chirality. Parity violation will serve as a powerful observable for distinguishing cosmological background GWs from astrophysical ones~\cite{Lue:1998mq}. Probing axion dark matter through future space-based gravitational-wave detectors will enable the exploration of broader parameter space for axions and ALPs. Except for the ground-based gravitational-wave observatories~\cite{Nagano:2019rbw, Heinze:2023nfb}, forthcoming space-based missions hold the potential to probe axion-like dark matter directly~\cite{Yao:2024fie, Gue:2024txz, Yao:2024hap}.

The SGWB arises from the superposition of GWs produced by a large number of independent sources~\cite{Christensen:2018iqi}. It exhibits stochasticity and has a signal strength that is relatively weak compared to the total intensity sensitivity of detectors, categorizing it as a weak signal, and methods for its detection have been developed~\cite{Allen:1997ad}. Due to the stochastic and uncorrelated nature of the general generation process, the SGWB is assumed to be unpolarized. However, parity violation in gravity, such as the Chern-Simons coupling ~\cite{Jackiw:2003pm,Alexander:2009tp} and the Nieh-Yan coupling in teleparallel equivalent of general relativity (TEGR)~\cite{Li:2020xjt,Li:2021wij,Cai:2021uup,Wu:2021ndf,Li:2023fto,Li:2022vtn,Rao:2023doc,Zhang:2024vfw,Xu:2024kwy}, can modify the generation and propagation of gravitational waves, leading to a circularly polarized SGWB. The chirality of GWs can be effectively measured within the frequency bands of several detectors, including ground-based detectors~\cite{Seto:2007tn,Seto:2008sr,Smith:2016jqs}, space-based instruments such as LISA~\cite{Seto:2006hf} and Taiji~\cite{Orlando:2020oko}, and through observations of the Cosmic Microwave Background (CMB)~\cite{Saito:2007kt,Sorbo:2011rz}.

LISA (Laser Interferometer Space Antenna) is a triangular GW detector in the orbit around the Sun, which is expected to be launched in the 2030s, with an arm length of $L=2.5 \times 10^9 $m~\cite{LISA:2017pwj}. Taiji is similar to LISA but has an arm length of $L=3  \times 10^9 $m~\cite{Ruan:2018tsw}. Due to their planar configuration, individual detectors are insensitive to the chiral signatures of GWs. For an isotropic SGWB, the detection of its circular polarization requires the correlation of two non-coplanar gravitational wave detectors~\cite{Seto:2005qy}. Therefore, a network of detectors is necessary, such as ground-based networks~\cite{LIGOScientific:2006zmq,Schutz:2011tw} or the space-based network LISA-Taiji~\cite{Seto:2020zxw,Ruan:2020smc,Orlando:2020oko,Wang:2021uih,Wang:2021njt,Cai:2023ywp,Zhao:2024yau, Chen:2024ikn}, which can enhance the detection of the circular polarization of the SGWB. Furthermore, space-based GW detector networks also provide numerous other advantages, including improved gravitational wave polarization measurements~\cite{Wang:2021mou}, enhanced parameter estimation for Galactic binaries~\cite{Zhang:2022wcp}, better sky localization accuracy~\cite{Zhang:2021wwd,Shuman:2021ruh}, more accurate localization of massive binaries~\cite{Ruan:2019tje}, detection of black hole formation mechanisms~\cite{Yang:2022cgm}, increased detection capabilities for stellar binary black holes~\cite{Chen:2021sco}, and increased precision of GW standard sirens and cosmological parameter estimation~\cite{Yang:2021qge,Wang:2021srv,Wang:2020dkc}.

To evaluate the detection capability of the LISA-Taiji network for chiral gravitational wave background (GWB), we estimate the spectral parameters and normalized model parameters of the chiral GWB generated by early cosmic axions using the Fisher information matrix. Additionally, we perform a Fisher analysis based on the fitted SGWB energy density spectrum from the dark photon coupling model~\cite{Machado:2019xuc} and broken power-law spectrum from the Nieh-Yan coupling model~\cite{Xu:2024kwy}.

The paper is organized as follows. In Sec.~\ref{audible axions and GW Spectrum}, we briefly introduce how axions generate chiral gravitational waves through coupling with dark photons or the Nieh-Yan term, and present the energy density spectrum of the resulting gravitational waves, along with fitted templates and parameters. In Sec.~\ref{sec:RespFunandSenCur}, we describe the configuration of the space-based GW detector network and calculate its response to GWs. In Sec.~\ref{sec:Fisher information matrix}, we derive the Fisher information matrix and determine the parameters for two different GW energy spectra with the network. In Sec.~\ref{conclusion}, we present the conclusion and discussion.
The calculations in this work are performed using the Python packages \texttt{numpy} and \texttt{scipy}, and the plots are generated using \texttt{matplotlib} and \texttt{GetDist}.

\section{Audible Axions and Chiral GW Background}
\label{audible axions and  GW Spectrum}

Several mechanisms that produce chiral GWs from audible axions have been explored in prior research. 
One is the coupling of axion to dark photon~\cite{Machado:2018nqk,Machado:2019xuc,Salehian:2020dsf}, while the other involves axion coupling to the parity-violating gravity such as Chern-Simons~\cite{Sun:2020gem,Li:2023vuu,Ding:2024} and Nieh-Yan modified gravity~\cite{Li:2020xjt,Li:2021wij,Cai:2021uup,Wu:2021ndf,Li:2023fto,Li:2022vtn,Rao:2023doc,Zhang:2024vfw,Xu:2024kwy}. 
The former just generates chiral GWs mediated by dark photons, while the latter can produce GWs directly and efficiently. 
In this section, we explore the GW spectrum template and fitting parameters produced by these mechanisms.

\subsection{Chiral GWB from Dark Photon coupling}
\label{subsec:fit_temp_DP}
The chiral GWB can be generated through the asymmetrical production of dark photons~\cite{Machado:2018nqk,Machado:2019xuc,Salehian:2020dsf}. In this mechanism, the total action can be expressed as
\begin{align}
    S_{\text{DP}} = \int \dd^{4}x \sqrt{-g}
\Big[&\frac{M_p^2}{2} R -\frac{1}{4}X_{\mu\nu}X^{\mu\nu}+ \frac{\alpha_{X} }{4\fx}  \phi X_{\mu\nu} \widetilde{X}^{\mu\nu}\nonumber\\
& -\frac{1}{2}\partial_{\mu} \phi \partial^{\mu}{\phi} - V(\phi) 
\Big].\label{DP_total_equation}
\end{align}
Here, $\fx$ is the decay constant of the axion, $\alpha_{X}$ is the coupling coefficient and $V(\phi)=m^{2}\fx^{2}\left[1-\cos\left(\frac{\phi}{{\fx}}\right)\right]$ is the
cosine-like potential with the axion mass $m$. The third term in this action leads to a nontrivial dispersion relation for the helicities of dark photons, which takes the form 
$ \omega_{X,\pm}^{2}=k_{X}^{2}\mp k_{X}\frac{\alpha_{X}}{\fx}\phi'$.
This indicates that the asymmetric production of dark photons results in an oscillating stress-energy distribution that sources gravitational waves.

Previous studies provide a well-fitting curve for the chiral gravitational wave energy density spectrum produced by dark photons. 
For the SGWB template, a suitable ansatz is~\cite{Machado:2019xuc}
\begin{align}
	\tilde{\Omega}_{\rm GW}(\tilde{f}_{p}) = \frac{\mathcal{A}_{s}\left(\tilde{f}_{p}/f_{s}\right)^p}{1+\left(\tilde{f}_{p}/f_{s}\right)^{p} \exp\left[\gamma (\tilde{f}_{p}/f_{s}-1)\right]},
	\label{eq:temp_AA}
\end{align}
where $\tilde{\Omega}_{\rm GW} \equiv \Omega_{\rm GW}(f) / \Omega_{\rm GW}(f_{p})$ represents the normalized GW energy density, $f_{p}$ denotes the peak frequency. $f$ denotes the GW frequency and $\tilde{f}_{p} \equiv f/f_{p}$ is the dimensionless normalized frequency. Moreover, $\mathcal{A}_{s}$, $f_{s}$, $\gamma$, $p$ are the fitting parameters. 

From the derivation in~\cite{Machado:2018nqk}, the peak amplitude and peak frequency of the GW spectrum, at the time of GW emission, are given by $f_{p}  \simeq (\alpha_{X}\theta)^{2/3} m \,, \hspace{2mm} \, 
\Omega_{\rm GW}(f_{p} ) \simeq \left( \frac{\fx}{M_P} \right)^4 \, \left( \frac{\theta^{2}}{\alpha_{X}}\right)^{\frac{4}{3}}  \,$.
Here, $\theta$ is the initial misalignment angle and $\Mp \simeq 2.4 \times 10^{18} \, \text{GeV}$ is the reduced Planck mass.
Considering the expansion of the Universe, which leads to redshifting, these quantities become~\cite{Machado:2018nqk}
\begin{align}\label{gauge boson fpeak0}
f_{p} ^{0}& \simeq (\alpha_{X}\theta)^{2/3} \, T_{0} \left( \frac{g_{s,0}}{g_{s,*}}\right)^{1/3} \left( \frac{m}{M_P}\right)^{1/2} \,,\\
\label{gauge boson Omega0}
\Omega_{\rm GW}^{0}(f_{p} ^{0}) &\simeq 1.67\times 10^{-4} g_{s,*}^{-1/3}\left( \frac{\fx}{M_P} \right)^4 \, \left( \frac{\theta^{2}}{\alpha_{X}}\right)^{\frac{4}{3}} \, .
\end{align}
Here, we choose the effective number of relativistic degrees of freedom $g_{s,*}=106.75$, because the mechanism occurs near the QCD phase transition. $ g_{s,0}=3.938 $  is the effective relativistic degree of freedom today when the temperature $ T_{0} =2.35  \times 10^{-13} $ GeV. Based on the equations above, to produce detectable GW signals within the mHz frequency band, we adopt the following parameter values: $m = 1.0 \times 10^{-2}$ eV, $\fx = 1.0 \times 10^{17}$ GeV, $\alpha_X = 55$, and $\theta = 1.2$, as proposed by~\cite{Machado:2018nqk}.

\subsection{Chiral GWB from Nieh-Yan coupling}
\label{subsec:fit_temp_NY}

The chiral GWB  can also be generated through an axion-like mechanism that couples to the Nieh-Yan term, resulting in the direct and efficient production of chiral GWB during the radiation-dominated epoch~\cite{Xu:2024kwy}. This generation arises from the tachyonic instability of gravitational perturbations induced by the Nieh-Yan term. 
The total action for this  mechanism can be written as
\begin{equation}\label{NY_total_equation}
\begin{split}
S_{\text{NY}} = \int \dd^{4}x \sqrt{-g}
\Big[& -\frac{M^{2}_{\text{p}}}{2}\hat{T}  + \frac{\alpha_{T}M^{2}_{\text{p}}}{4{\fh}} \phi \hat{T}_{A\mu\nu} \widetilde{T}^{A\mu\nu}  \\
& -\frac{1}{2}\partial_{\mu} \phi \partial^{\mu}{\phi} - V(\phi)
\Big].
\end{split}
\end{equation}
Here, $\hat{T}$ is the torsion scalar, which is dynamically equivalent to the Ricci scalar in general relativity. Similar to the action in Eq.\eqref{DP_total_equation},
$\fh$ is the axion decay constant, $\alpha_{T}$ is the coupling coefficient and $V(\phi)$ represents the cosine-like potential.
The second term in Eq.\eqref{NY_total_equation} can also lead to a non-trivial dispersion relation for the GW helicities, given by
$\omega_{T,\pm}^{2}=k_{T}^{2} \pm\frac{\alpha_{T} \phi'}{{\fh}\Mp^{2}}k_{T}$. The $k_{T}$ here is the wave vector for GWs, indicating that the last term produces an effect analogous to the term $\frac{\alpha_{X} }{4{\fx}}  \phi X_{\mu\nu} \widetilde{X}^{\mu\nu}$ in the dark photon case. 

Specifically, when the axion field oscillates, one of the GW helicities will have a range of modes with imaginary frequencies, resulting in a tachyonic instability that drives exponential growth. Since the growth rate is related to helicities, the left-handed and right-handed GWs are generated asymmetrically, ultimately leading to chiral GWB. In~\cite{Xu:2024kwy}, the broken power-law template provides a better fit for the GW spectrum in this model, which can be written as
\begin{equation}\label{BPL template}
    \tilde{\Omega}_{T} = \left(\tilde{f}_{c}\right)^{\alpha_{1}}\left[1+0.75\left(\tilde{f}_{c}\right)^{\Delta}\right]^{\frac{(\alpha_{2}-\alpha_{1})}{\Delta}}.
\end{equation}
Here, $\tilde{\Omega}_{T} \equiv \Omega_{T}(f)/\Omega_{c}$, where $\Omega_{c}$ is the characteristic energy density. $\tilde{f}_{c} \equiv f/f_{c}$ is the dimensionless normalized frequency with the characteristic frequency $f_{c}$. Moreover, $\alpha_1$, $\alpha_2$ and $\Delta$ are fitting parameters.

\begin{figure}[t]
  \centering
  \captionsetup{justification=raggedright,singlelinecheck=false} 
  \includegraphics[width=0.95\columnwidth]{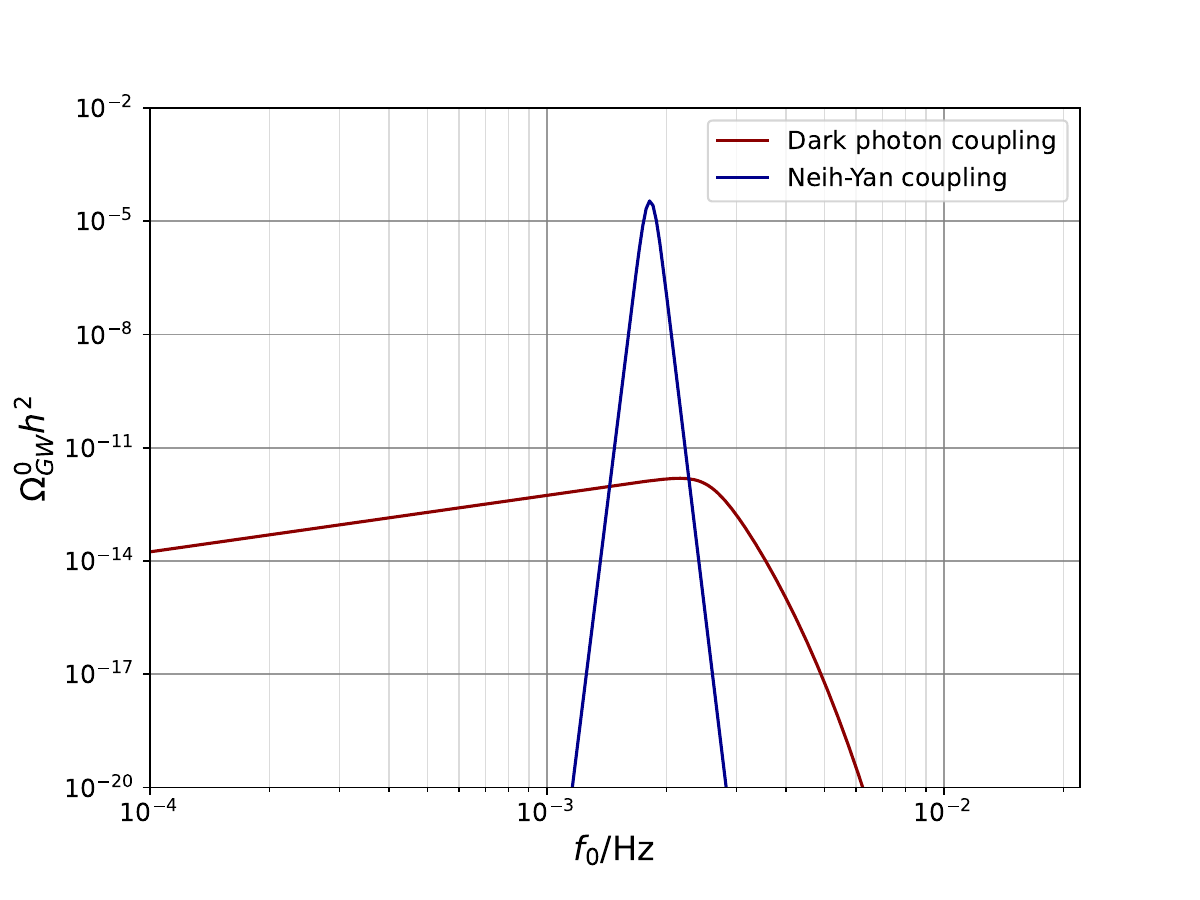} 
\caption{The broken power-law fitted curves of the SGWB for the dark photon (red) and Nieh-Yan (blue) coupling models, each using two parameter sets as described in Sec.~\ref{subsec:fit_temp_DP} and~\ref{subsec:fit_temp_NY}.}
  \label{fig:EnergyDensity curve} 
\end{figure}

\begin{figure*}[htbp]
  \centering
  \captionsetup{justification=raggedright,singlelinecheck=false} 
  \includegraphics[width=0.80\textwidth]{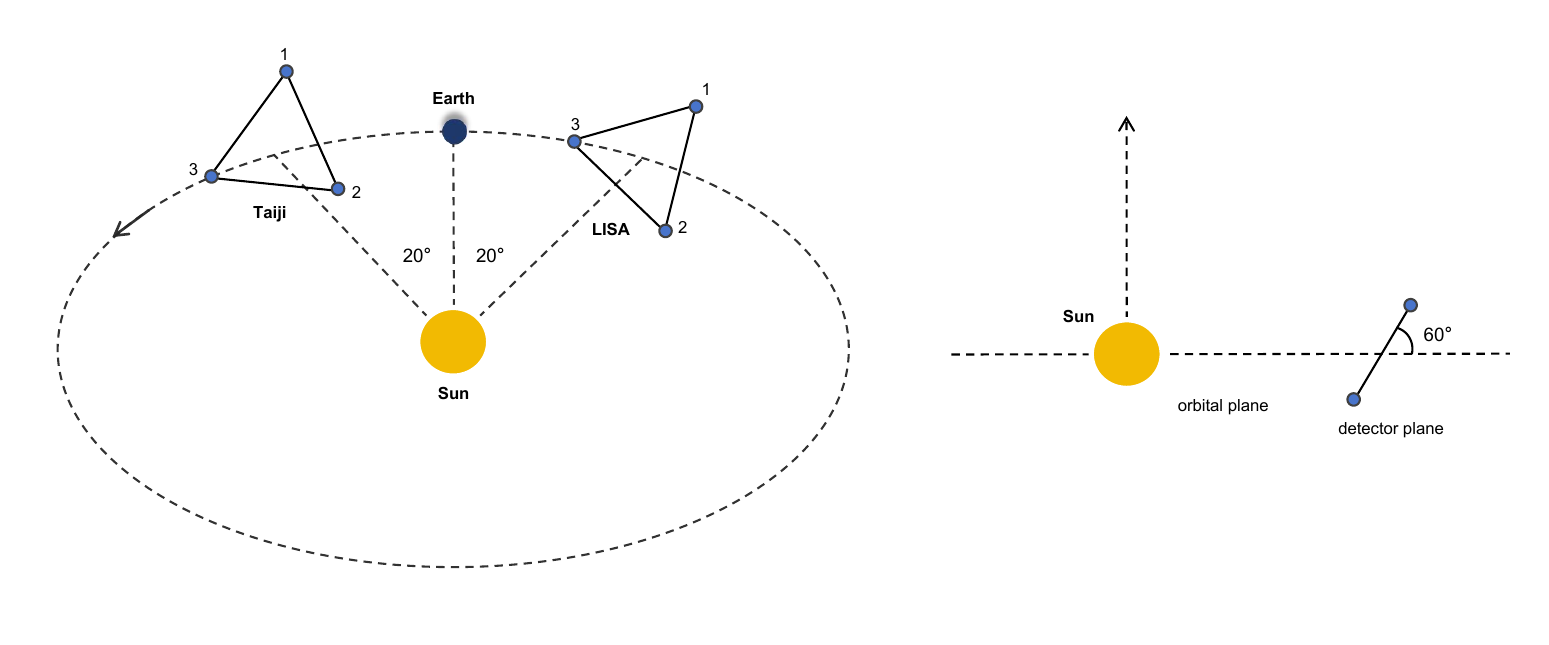} 
  \caption{The configuration of the LISA-Taiji joint network, including the spacecraft numbering scheme. LISA orbits 20 degrees behind the Earth, while Taiji precedes the Earth by the same angle. Both detector planes are inclined at 60 degrees relative to the ecliptic plane.}
  \label{fig:Lisa-Taiji_network_configuration}
\end{figure*}

In this equation, the characteristic frequency today, $f_{c}^{0}$, can be expressed in terms of physical parameters as
\begin{equation}\label{Nieh-Yan fc}
f_{c}^{0} = 0.7125 \text{ mHz} \left(\frac{100}{g_{s,*}}\right)^{\frac{1}{12}} \left(\frac{9}{14}\alpha_{T}\theta \right)^{\frac{2}{3}}\left(\frac{m}{\text{eV}}\right)^{\frac{1}{2}}.
\end{equation}
Here, we also choose $g_{s,*}=106.75$, as this mechanism occurs near the QCD phase transition. 
The characteristic energy density $\Omega_{c}^{0}$  can similarly be written in terms of physical parameters as
\begin{equation}\label{Nieh-Yan Omegac}
    \Omega_{c}^{0} = \frac{\theta^{2}\fh^{2}/2}{3\Mp^{2}} \frac{m^{2}}{H_{\text{osc}}^{2}}\simeq \left(\frac{\theta {\fh}}{\Mp}\right)^{2}.
\end{equation}
For the Nieh-Yan coupling model, we choose the following parameters to generate detectable gravitational wave signals in the mHz band: $m = 0.1$ eV, $\fh = 1.0 \times 10^{17}$ GeV, $\alpha_T = 35.61$, and $\theta = 1$~\cite{Xu:2024kwy}.

In Fig.~\ref{fig:EnergyDensity curve}, we present the broken power-law fit curves of the SGWB spectrum generated by the dark photon coupling model and the Nieh-Yan coupling model.

\section{Network of Space-based GW detectors}
\label{sec:RespFunandSenCur}

In this section, we adopt the commonly used orbits of LISA and Taiji, combining them to evaluate their effectiveness in detecting the SGWB.
We establish the coordinate system in the Solar System Barycentric Coordinate System (SSB).
LISA trails the Earth by 20 degrees, and Taiji leads by the same degree, with both detector planes tilted 60 degrees relative to the ecliptic plane.
The LISA-Taiji network configuration is displayed as Fig.~\ref{fig:Lisa-Taiji_network_configuration}.

\begin{figure*}[htbp]
  \centering
  \captionsetup{justification=raggedright,singlelinecheck=false} 
  \begin{subfigure}[b]{0.45\textwidth}
    \centering
\includegraphics[width=\textwidth]{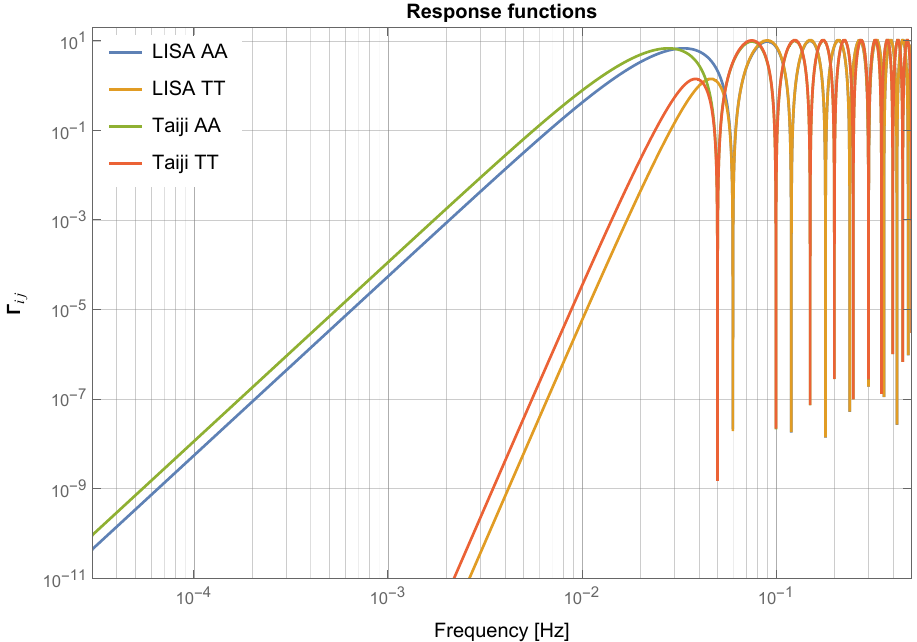}
\caption{The self-correlation response functions $\Gamma_{ij}(f)$.}
\label{fig:Response_functions}
\end{subfigure}
  \hspace{0.02\textwidth}
  \begin{subfigure}[b]{0.45\textwidth}
    \centering
\includegraphics[width=\textwidth]{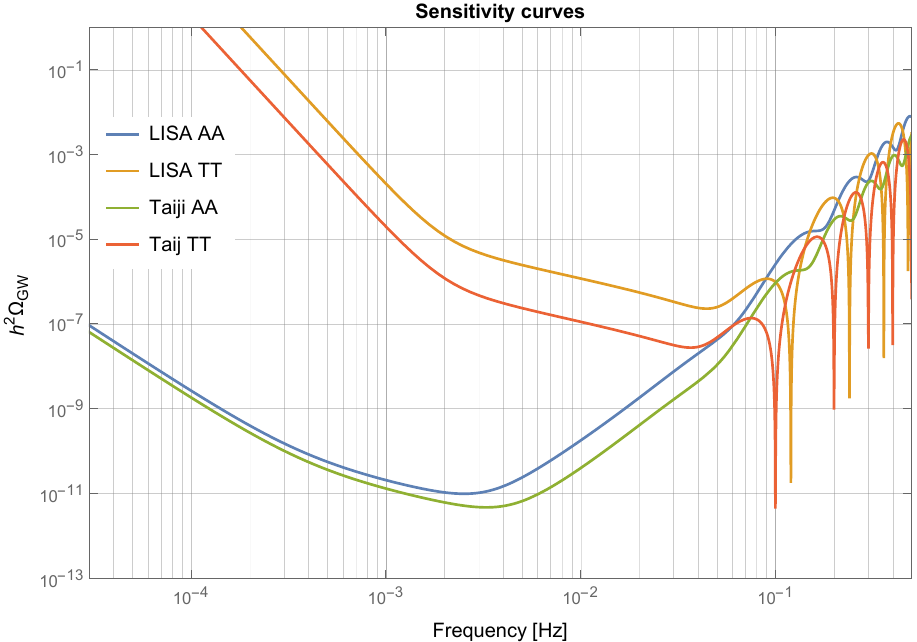}
\caption{Sensitivity curves for the total intensity of GWs.}
\label{fig:LISA_strain_sensitivities}
  \end{subfigure}
  \caption{(a) The self-correlation response functions $\Gamma_{ij}$ in~\eqref{SelfResponseFunction} of a single detector, for the respective TDI channels of LISA and Taiji, where the E channel has the same result as the A channel. 
(b) 
The sensitivity curves for the total intensity of GWs in Eq.~\eqref{NoiseEnergyDensity} for the self-correlation of the respective channels of LISA and Taiji, where again the E-channel has the same results as the A-channel.}
\label{fig:combined_response_LISA}
\end{figure*}

\subsection{Noise and sensitivity of the detectors}

Each detector contains three interferometers that simultaneously detect the Doppler shift induced by GWs. 
The data stream of Time-Delay Interferometry (TDI) channel $i$ is given by
\begin{equation}
    d_{i} (t) = s_{i} (t) + n_{i} (t),
\end{equation}
where $s_{i} (t)$ represents the signal and  $n_{i} (t)$ denotes the instrumental noise.
In general, it is more convenient to work in the frequency domain
\begin{equation}
    \tilde{d}_{i}  (f) = \int_{-T/2}^{T/2} \dd t \; e^{2 \pi i f t} d_{i} (t),
\end{equation}
where $T$ represents the observation time.
In this paper, we assume that the noise is Gaussian and uncorrelated.
The respective correlations of the signal and noise in the frequency domain can be expressed as
\begin{equation}
    \begin{aligned}
\left\langle\tilde{s}_i(f) \tilde{s}_j^*\left(f^{\prime}\right)\right\rangle & =\frac{1}{2} S_{i j}(f) \delta\left(f-f^{\prime}\right) ,
\\
\left\langle\tilde{n}_i(f) \tilde{n}_j^*\left(f^{\prime}\right)\right\rangle & =\frac{1}{2} N_{i}(f)\delta_{ij} \delta\left(f-f^{\prime}\right),
\end{aligned}
\end{equation}  
where $S_{ij} (f)$ and $N_{i}(f)$ are the one-sided signal and noise power spectral density (PSD), respectively. 
Assuming independent TDI channel noises (e.g., in the A, E, T combination, which are the optimal TDI variables for LISA-like detectors), $S_{ij} (f)$ can be expressed as
\begin{equation}
\label{eq:psd}
\begin{aligned}
& S_{ij}(f) = \sum_\lambda P_\lambda(f) \Gamma_{ij}^\lambda(f) = 
     \sum_\lambda P_\lambda(f) \times
    \\
&\left[ 
    \left(2 \pi k L_i\right) \left(2 \pi k L_j\right) W\left(k L_i\right) W^*\left(k L_j\right) \tilde{\Gamma}_{ij}^\lambda(f) + \text{h.c.}
    \right] .
\end{aligned}
\end{equation}
Here, $k =f/c$,  $ \lambda = L$ or $ R $ identifies left- and right-handed polarizations, 
$L_i$ and $L_j$ are the detector armlengths,
$ \Gamma_{ij}^\lambda(f) $ is the full detector response function and $P_\lambda(f) $ is the  GW power spectrum. 
The function $W(kL)$ represents the phase delay due to the detector arm length, as detailed in Appendix~\ref{appendix:Response_functions}. 
$\tilde{\Gamma}_{i j}^\lambda(k)$ denotes the geometrical contribution to the detector response function for the correlation between channels $i$ and $j$, as detailed in equation~\eqref{GeoResponseFunction}.

In this work, we adopt the standard two-parameter noise model used for LISA, which accounts for the two dominant noise sources in space-based GW detectors: acceleration (acc) noise and Optical Measurement System (OMS) noise. For Taiji, we use a similar noise model with distinct parameters $A_{\rm acc}$ and $A_{\rm OMS}$.
The acceleration noise power spectrum $P_{\text{acc}}(f)$ and OMS noise power spectrum $P_{\text{OMS}}(f)$ are given by~\cite{wgLISADataChallenge2020,Armano:2016bkm}
\begin{align}
P_{\mathrm{acc}}(f) & =A_{\rm acc}^2  \left[1+\left(\frac{0.4 \mathrm{mHz}}{f}\right)^2\right]\left(\frac{2 \pi f}{c}\right)^2
\\
&\quad \qquad  \times  \left[1+\left(\frac{f}{8 \mathrm{mHz}}\right)^4\right]\left(\frac{1}{2 \pi f}\right)^4,\nonumber\\
P_{\mathrm{OMS}}(f) & =A_{\rm OMS}^2 \left[1+\left(\frac{2 \mathrm{mHz}}{f}\right)^4\right]\left(\frac{2 \pi f}{c}\right)^2.
\end{align}
Here, $A_{\rm acc}$ and $A_{\rm OMS}$ are the amplitudes of the acceleration noise and the OMS noise, respectively. 
For the two detectors, the noise amplitude parameters for LISA~\cite{Babak:2021mhe} and Taiji~\cite{Luo:2019zal,Luo:2021qji,Ruan:2020smc} are listed in Table~\ref{table_detectors}.
\begin{table}[h]
\centering
\begin{tabular}{|c|c|c|c|}
\hline
 & $A_{\rm OMS}$ & $A_{\rm acc}$ & $L$  \\ \hline
LISA & $15\, {\rm pm}/\sqrt{{\rm Hz}}$ & $3\,{\rm fm/s^2}/\sqrt{{\rm Hz}}$ &  $2.5\, {\rm Gm}$  \\ \hline
Taiji  & $8\,{\rm pm}/\sqrt{{\rm Hz}}$  &  $3\,{\rm fm/s^2}/\sqrt{{\rm Hz}}$ &  $3.0\, {\rm Gm}$  \\ \hline
\end{tabular}
\caption{Noise amplitude spectral density parameters and arm lengths for different space-based GW detectors.}
\label{table_detectors}
\end{table}

For convenience, we define the detector's characteristic frequency as $f_{\ast} \equiv c/2\pi L$. 
The power spectral density of the noise for a single detector channel is then given by~\cite{Prince:2002hp,Flauger:2020qyi}
\begin{equation}
\begin{aligned}
& N_{\mathrm{A}}(f) =  N_{\mathrm{E}}(f) \\
= & 8  \sin ^2\left(\frac{f}{f_{\ast}}\right)\left\{4\left[1+\cos \left(\frac{f}{f_{\ast}}\right)+\cos ^2\left(\frac{f}{f_{\ast}}\right)\right] \times\right. 
\\&\qquad\quad P_{\rm acc}(f)  \left.+\left[2+\cos \left(\frac{f}{f_{\ast}}\right)\right] \times P_{\rm OMS}(f) \right\},
\end{aligned}
\end{equation}
and
\begin{equation}
\begin{aligned}
N_{\mathrm{T}}(f) =& 
16 \sin ^2\left(\frac{f}{f_{\ast}}\right)\left\{2\left[1-\cos \left(\frac{f}{f_{\ast}}\right)\right]^2 \times  \right. 
\\
& \left.  P_{\rm acc}(f)+\left[1-\cos \left(\frac{f}{f_{\ast}}\right)\right] \times P_{\rm OMS}(f) \right\}
,
\end{aligned}
\end{equation}
where the subscripts A, E, and T denote 
the noise-orthogonal TDI channels A, E, and T, respectively.

To directly compare incident GW signals with detector noise, we define the strain sensitivity of total intensity for all GW modes as~\cite{Flauger:2020qyi}
\begin{equation}
P_{N,i i}(f)=\frac{N_{i }(f)}{\Gamma_{ii}^{}(f)}\,
    ,
\end{equation}
where $\Gamma_{ii}^{}(f)$ represents the sky-averaged response functions of individual TDI channels, which can be calculated via Eq.~\eqref{SelfResponseFunction}.
The corresponding total intensity, in GW energy density units, is given by:
\begin{equation}\label{NoiseEnergyDensity}
    h^2 \Omega_{N, i i}(f) = \frac{4 \pi^2 f^3}{3\left(H_0 / h\right)^2} P_{N, i i}(f) 
    \;,
\end{equation}
where $H_0=h \, 100 \, \text{km} \, \text{s}^{-1} \, \text{Mpc}^{-1}$ is the value of the present-day Hubble parameter and $h=0.67 $ is the dimensionless Hubble parameter. 
Fig.~\ref{fig:LISA_strain_sensitivities} shows the total intensity sensitivity curves for the A and E channels of both LISA and Taiji.
Similarly, the SGWB adopts a similar notation, with $P_{N,ij}$ replaced by the signal PSD~\cite{Orlando:2020oko}
\begin{equation}\label{EnergyDensity}
    h^{2}\Omega_{\mathrm{GW}}^{\lambda}(f)=\frac{4\pi^{2}f^{3}}{3(H_{0}/h)^{2}}P_{\lambda}(f)
    \,.
\end{equation}

\begin{figure*}[htbp]
  \centering
  \captionsetup{justification=raggedright,singlelinecheck=false} 
  \begin{subfigure}[b]{0.9\columnwidth}
    \centering
    \includegraphics[width=\textwidth]{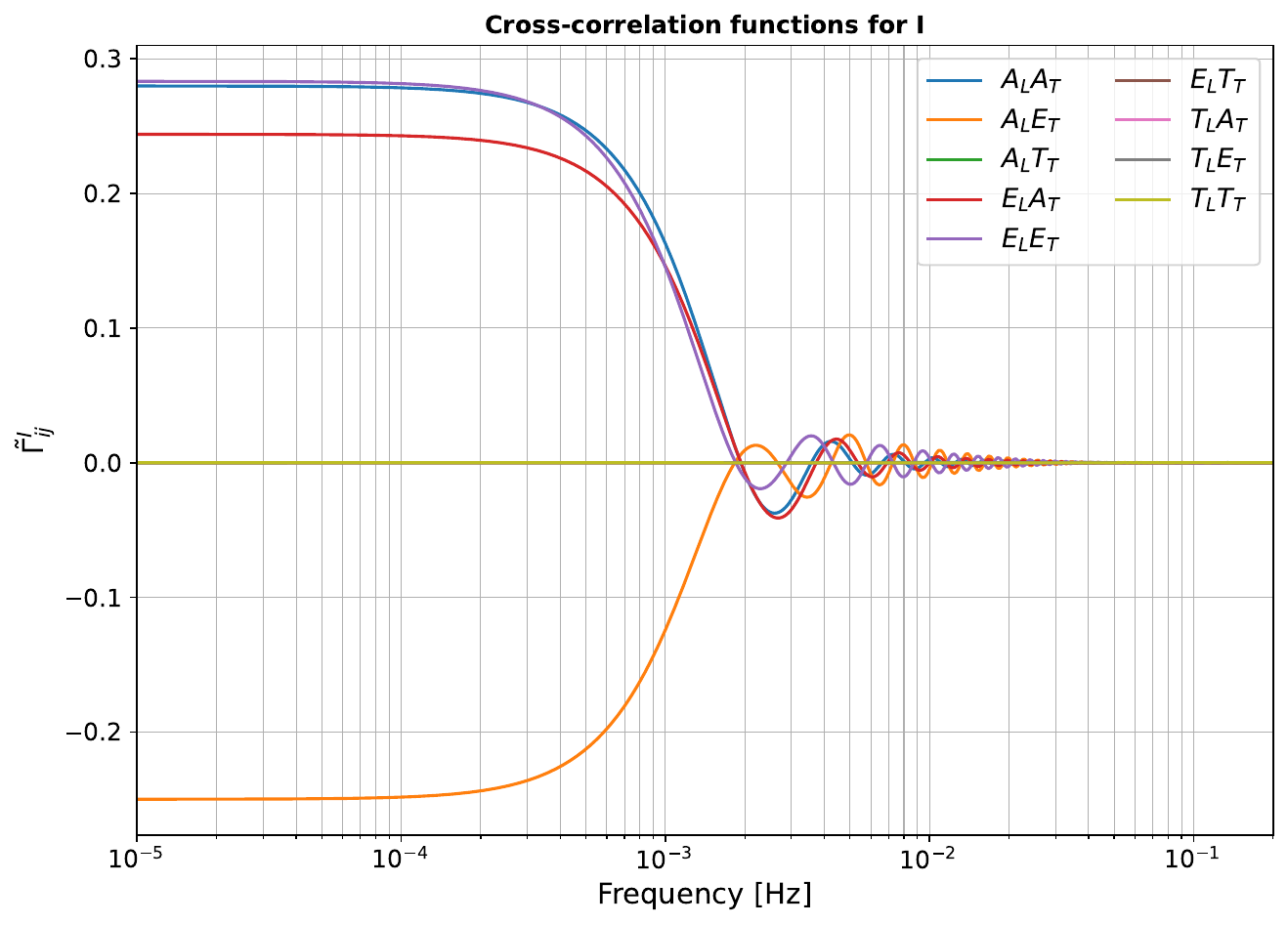}
    \caption{Cross-correlation functions $\tilde{\Gamma}^I_{ij}(f)$  for LISA-Taiji.}
    \label{fig:I_parameter}
  \end{subfigure}
  \hspace{0.00\textwidth}   
  \vspace{-0.0cm} 
  \begin{subfigure}[b]{0.9\columnwidth}
    \centering
    \includegraphics[width=\textwidth]{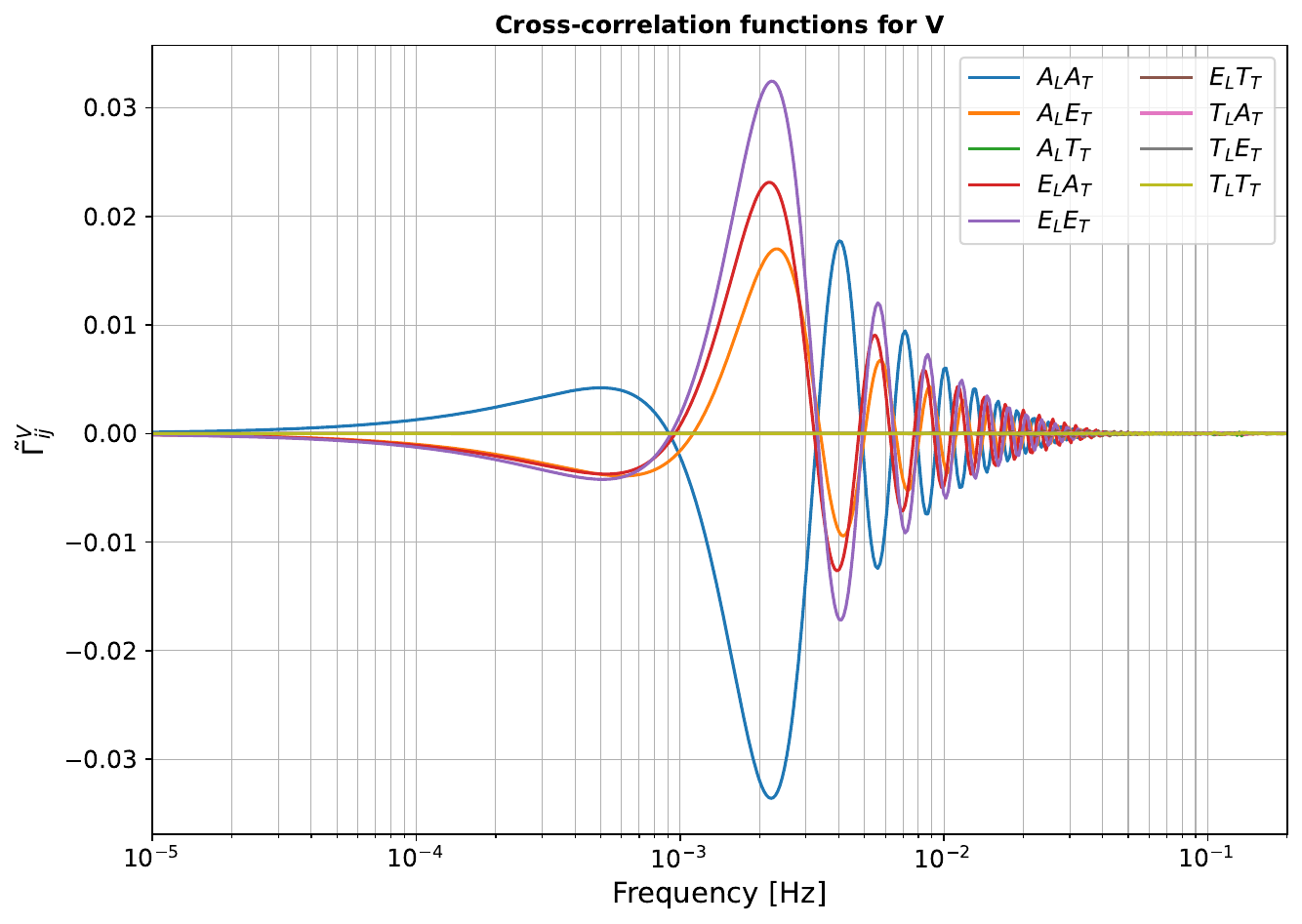}
\caption{Cross-correlation functions $\tilde{\Gamma}^V_{ij}(f)$  for LISA-Taiji.}
    \label{fig:V_parameter}
  \end{subfigure}
  \caption{
  Cross-correlation functions  $\tilde{\Gamma}^{\lambda}_{ij}(f)$ in Eq.~\eqref{GeoResponseFunction} between the TDI channels of LISA and Taiji, for Stokes parameter $I$ in (a) and $V$ in (b).
}
  \label{fig:The geometric factor of the response function}
\end{figure*}

\subsection{LISA-Taiji cross-correlations}

We use the Stokes parameters $I(f)$ and $V(f)$ to characterize the polarization of the SGWB in the cross-correlated detector data stream.
They are defined as
\begin{equation}
I(f)=P_R(f)+P_L(f), \quad 
V(f)=P_R(f)-P_L(f) .
\label{Stokes-parameter}
\end{equation}
Here, $I$ represents the total intensity of the GW, while $V$ quantifies the difference between right-handed and left-handed circular polarization intensities. 
Parity-violating effects in the early Universe may give rise to a nonzero value of $V$. 
By using detectors to measure it, we can extract information about the circular polarization of GWs.
We can express GW power spectral density $ S_{ij} $ as 
\begin{equation}
     S_{ij}(f) = I(f)  \Gamma_{ij}^{I}(f) + V(f) \Gamma_{ij}^{V}(f),
\end{equation}
where $\Gamma_{ij}^{I}(f)$ and $\Gamma_{ij}^{V}(f)$ are the overlap reduction functions for intensity and circular polarization components, quantifying the correlated response between TDI channels $i$ and $j$.
These functions are defined as:
\begin{equation}\label{IV_Response_Function}
\begin{aligned}
    \Gamma ^I_{ij}(f) &= \frac{\Gamma ^R_{ij}(f) + \Gamma ^L_{ij}(f)}{2}, \\
\Gamma ^V_{ij}(f) &= \frac{\Gamma ^R_{ij}(f) - \Gamma ^L_{ij}(f)}{2}.
\end{aligned}
\end{equation}
Thus, the power spectral density in Eq.~\eqref{eq:psd} for $\tilde{\Gamma}_{ij}^\lambda(f)$ can be reformulated using the Stokes parameters $I$ and $V$, with $\lambda = I, V$.
By cross-correlating the signals from LISA and Taiji channels, we can extract nonzero $ \tilde{\Gamma} ^{V}_{ij} (f) $ values. The $I$ and $V$ components resulting from this cross-correlation of all TDI channels between LISA and Taiji are presented in Fig.~\ref{fig:The geometric factor of the response function}.

Additionally, we introduce the circular polarization parameter as
\begin{equation}\label{CircularPolarizationParameter}
    \Pi(f) = \frac{V(f)}{I(f)} 
    .
\end{equation}
The correlation between the outputs of different detectors can be expressed as:
\begin{equation}\label{CrossCorelation}
\left\langle\mathcal{C}_{ij}\right\rangle = \left\langle \tilde{d}_{i} \tilde{d}_{j} \right\rangle = \frac{1}{2}\left[\Gamma_{ij}^I(f) I(f) + \Gamma_{ij}^V(f) V(f)\right].
\end{equation}

Assuming that the noise is Gaussian, the likelihood function of the signal model is~\cite{Chen:2024fto}
\begin{equation}\label{likelihood function}
\begin{aligned}
&    \mathcal{L} = p(\mathcal{C} \mid \theta ) \\  
    & \propto \exp \left\{-\frac{T_{\mathrm{obs}}}{2} \sum_\kappa \int_0^{\infty} \dd f \frac{\left[2 \mathcal{C}_\kappa-\left(\Gamma_\kappa^I I+\Gamma_\kappa^V V\right)\right]^2}{N_\kappa^2(f)} \right\} ,
\end{aligned}
\end{equation}
where $\kappa = \{A_{L}-A_{T}, A_{L}-E_{T}, E_{L}-A_{T}, E_{L}-E_{T}\}$ represent the independent channel pairs of LISA and Taiji. With $A_{L}$ and $E_{L}$ denoting the LISA channels and $A_{T}$ and $E_{T}$ corresponding to the Taiji channels. 
$T_{\mathrm{obs}}$ denotes the effective observation time, which is set to 3 years in this work. 
The noise term $N_\kappa(f)$ is defined as $N_\kappa(f) = \sqrt{N_{i} \left ( f \right ) N_{j} \left ( f \right ) }$. 
For strong GW signal, such as an SGWB with a large signal-to-noise ratio (SNR), $N_\kappa^2(f)$ in~\eqref{likelihood function} is replaced by~\cite{Cornish:2001qi,Cornish:2001bb,Chen:2024fto}
\begin{equation}\label{strong signals}
    M_{ij}(f) = (N_i + \Gamma_{ii}^{} I)(N_j + \Gamma_{jj}^{} I) + \left( \Gamma_{ij}^{I} I + \Gamma_{ij}^{V} V \right)^2
    .
\end{equation}

\section{Fisher Matrix Analysis}
\label{sec:Fisher information matrix}

In this section, we employ Fisher matrix analysis to estimate the measurement accuracy of the GW spectral parameters.
The Fisher matrix is given by as~\cite{Orlando:2020oko,Chen:2024fto}
\begin{equation}
    F_{a b}  = - \sum_\kappa 4 T_{\text {obs }} \int_0^{\infty} \dd f \frac{\frac{\partial\left\langle C_\kappa\right\rangle}{\partial \theta_a} \frac{\partial\left\langle C_\kappa\right\rangle}{\partial \theta_b}}{N_\kappa^2(f)},
\end{equation}
where $\theta_a$ and $\theta_b$ are the model parameters. 
The term $C_\kappa$ is the correlation of the observed data between the $\kappa$ channel sets, and $N_\kappa(f)$ represents the signal variance caused by noise.
For the frequency integration, we take the lower cutoff at $ 10^{-5} $ Hz and the upper cutoff at $ 10^{-1} $ Hz. In this study, we assume a frequency-independent circular polarization parameter $\Pi(f) = \Pi$ and derive the Fisher matrix expression for the GW model parameters as follows.

By substituting the signal Eq.~\eqref{EnergyDensity}, the circular polarization parameter in Eq.~\eqref{CircularPolarizationParameter}, and Eq.~\eqref{CrossCorelation}, we have
\begin{equation}
\begin{aligned}
{F}_{a b} = & \,4 T_{\rm obs} \left (  \frac{3 H_{0}^{2}}{4 \pi^{2}  }    \right ) ^{2} \times \\
& \sum_{\kappa }^{}  \int_{0}^{\infty } \dd f  
\frac{\left ( \Gamma_{\kappa }^{I} + \Pi \,\Gamma_{\kappa }^{V}  \right )^{2} \partial_{\theta_{a}  }  \Omega \left ( f \right ) 
 \partial_{\theta_{b}  }  \Omega \left ( f \right )
   }
{f^{6} \; N_{\kappa }^{2} },
\end{aligned}
\end{equation}
with $a$ and $b$ indicate both parameters of the GW model.
For example, when one parameter is $\Pi$ and the other is a GW model parameter
\begin{equation}
\begin{aligned}
{F}_{a \small \Pi} = & \,4 T_{\rm obs} \left (  \frac{3 H_{0}^{2}}{4 \pi^{2}  }    \right ) ^{2} \times \\
& \sum_{\kappa }^{}  \int_{0}^{\infty } \dd f  
\frac{\Gamma_{\kappa }^{V} \left ( \Gamma_{\kappa }^{I} + \Pi \,\Gamma_{\kappa }^{V}  \right ) \,  \Omega \left ( f \right ) 
 \partial_{\theta_{a}  }  \Omega \left ( f \right )    }
{f^{6} \; N_{\kappa }^{2} } .
\end{aligned}
\end{equation}
When both parameters in the Fisher matrix are $\Pi$
\begin{equation}
\begin{aligned}
{F}_{\small \Pi \small \Pi} =  4 T_{\rm obs} \left (\frac{3 H_{0}^{2}}{4 \pi^{2}  }    \right ) ^{2}
 \sum_{\kappa }^{}  \int_{0}^{\infty } \dd f  
\frac{ \left ( \Gamma_{\kappa }^{V} \right )^{2} \Omega \left ( f \right )^{2} }{f^{6} \; N_{\kappa }^{2} } .
\end{aligned}
\end{equation}

\subsection{GW spectral parameters}

\begin{figure*}[ht]
  \centering
  \captionsetup{justification=raggedright,singlelinecheck=false}
  \includegraphics[width=0.80\textwidth]{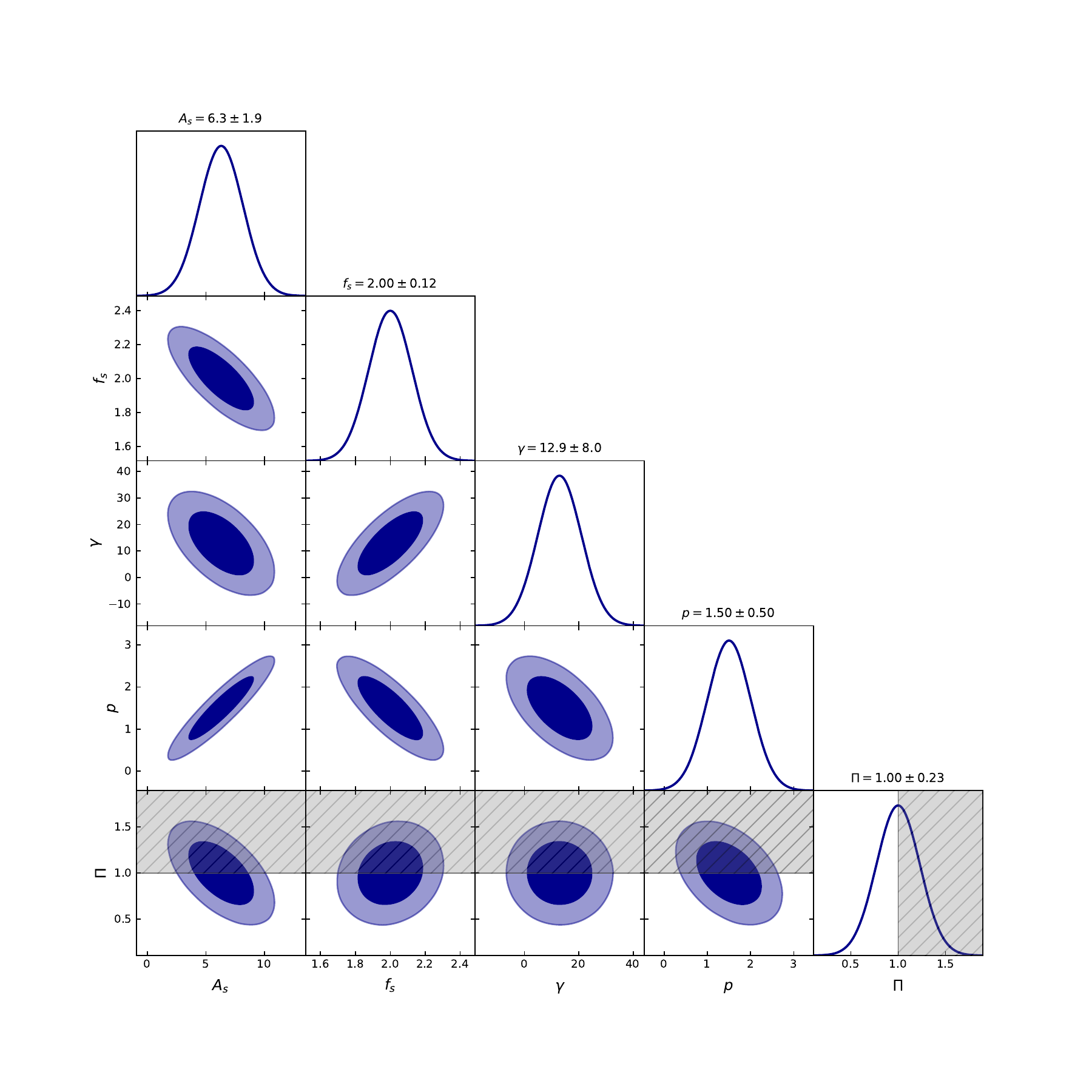} 
\caption{Corner plots of SGWB spectral parameters estimates for the dark photon coupling model derived from the Fisher matrix, with parameter values listed in Table~\ref{tab:parameter_values_AA}. At the top of each column, the corresponding parameters' 1$\sigma$ uncertainty are presented. The gray shaded areas correspond to regions of the parameter space with $\Pi > 1$, which is theoretically unacceptable. }
  \label{fig:corner_plots_AudibleAxio} 
\end{figure*}

\begin{figure*}[htbp]
  \centering
  \captionsetup{justification=raggedright,singlelinecheck=false} 
  \includegraphics[width=0.90\textwidth]{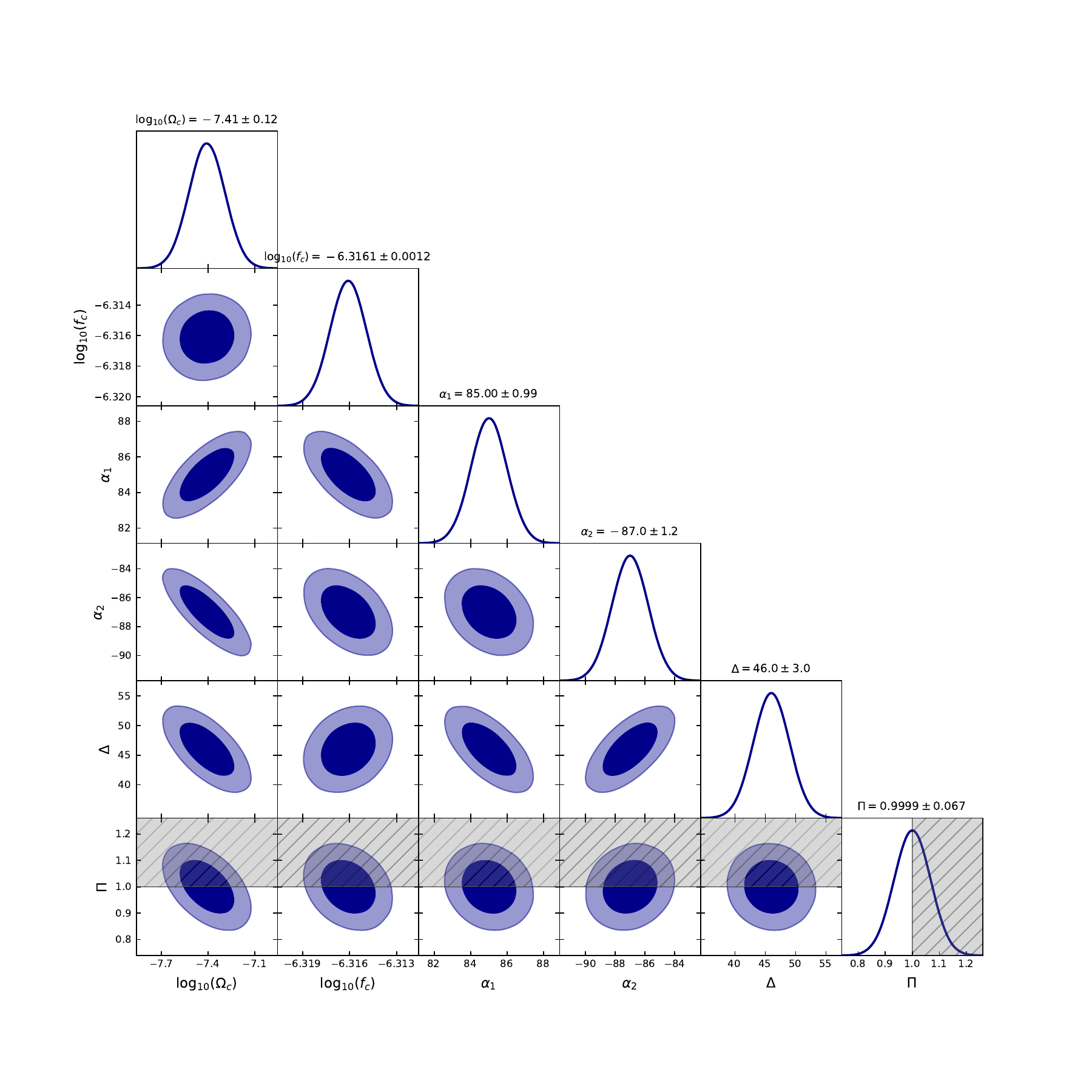} 
  \caption{Corner plots of SGWB spectral parameters estimates for the Nieh-Yan coupling model derived from the Fisher matrix, with parameter values listed in Table~\ref{tab:parameter_values_BPL}. 
  At the top of each column, the corresponding parameters' 1$\sigma$ uncertainty is presented. 
  The gray shaded areas correspond to regions of the parameter space with $\Pi > 1$, which is theoretically unacceptable. 
  }
  \label{Corner plots BPL} 
\end{figure*}

We selected the spectral parameters of the GW template for parameter estimation, as detailed in Table~\ref{tab:parameter_values_AA}, and the results from the Fisher analysis of the GW energy density spectrum template~\eqref{eq:temp_AA} for audible axion are illustrated in Fig.~\ref{fig:corner_plots_AudibleAxio}. 

\begin{table}[H]
  \centering
  \begin{tabular}{|c|c|c|c|c|c|}  
    \hline
    $A_s$ & $f_s$ & $\gamma$ & $p$ & $\Pi$ \\ \hline
    $6.3$ & $2.0$ & $12.9$ & $1.5$ & $ 0.9999 $ \\ \hline
  \end{tabular}
  \caption{Parameter values for the broken power-law template~\eqref{eq:temp_AA} for the dark photon coupling model.}
  \label{tab:parameter_values_AA}
\end{table}

For the dark photon coupling model, the uncertainties in the parameter estimates are illustrated graphically, encompassing four spectral parameters $\left\{A_s, f_s, \gamma, p\right\}$ and the circular polarization parameter $\Pi$. 
The confidence ellipses indicate that the true parameter values lie within the inner ellipse at a $1\sigma$ confidence level and within the outer ellipse at a $2\sigma$ confidence level. 
At the $1\sigma$ confidence level, the relative errors are less than $62.0\%$ for spectral parameters and less than $23.0\%$ for $\Pi$.
The elongation of the confidence ellipses reflects the strength of the correlation among spectral parameters, with more elongated ellipses indicating stronger correlations. 
Specifically, $p$ and $A_s$ exhibit a strong correlation. Additionally, the $\Pi$ is negatively correlated with $A_s, p$, shows a weak negative correlation with $\gamma$ and is independent of $f_s$. 
In Figs.~\ref{fig:corner_plots_AudibleAxio},~\ref{Corner plots BPL},~\ref{fig:corner_plots_AudibleAxio_Fundamentalparameters}, and~\ref{Corner plots BPL_Fundamentalparameters}, the gray solid line represents the fiducial value $\Pi = 1$, representing a fully polarized state. 
The gray shaded areas in the corner plots correspond to regions of the parameter space where $\Pi > 1$, which is theoretically unacceptable.

For the broken power-law template in the Nieh-Yan coupling models, we also have selected the spectral parameters as shown in Table~\ref{tab:parameter_values_BPL}, with the corresponding Fisher analysis results presented in Fig.~\ref{Corner plots BPL}.
The direct coupling of the axion to the gravitational field in Eq.\eqref{NY_total_equation} and the parameter choices in Table \ref{tab:parameter_values_BPL} result in a strong signal, making the variance assumption invalid. Therefore, the noise term $ N_{\kappa }^{2} $ in our Fisher matrix is replaced by $ M_{\kappa}(f)$ in Eq.~\eqref{strong signals}. The Fisher analysis results for the GW energy density spectrum in the Nieh-Yan coupling model are shown in Fig.~\ref{Corner plots BPL}. At the $1\sigma$ confidence level, the relative errors are less than $31.8\%$ for spectral parameters and less than $6.7\%$ for $\Pi$. $\Omega_{\mathrm{c}}$ and $f_{\mathrm{c}}$ exhibit relatively independent errors, while other spectral parameters show significant correlations. Furthermore, the circular polarization parameter $\Pi$ exhibits a negative correlation with $\Omega_{\mathrm{c}}$ and $f_{\mathrm{c}}$, a statistically weak correlation with $\alpha_1$ and $\alpha_2$, and no correlation with $\Delta$.

\begin{table}[H]
  \centering
  \begin{tabular}{|c|c|c|c|c|c|}  
    \hline
    $\Omega_{c}$ & $f_c   $/mHz & $\alpha_1$ & $\alpha_2$ & $\Delta$ & $\Pi$ \\ \hline
    $6.072 \times 10^{-4}$ & $1.807 $ & 85 & -87 & 46 & 0.9999 \\ \hline
  \end{tabular}
  \caption{Parameter values for the broken power-law template~\eqref{BPL template} for the Nieh-Yan coupling model.}
  \label{tab:parameter_values_BPL}
\end{table}

\subsection{Normalized model parameters}

Based on the relationship between the spectral  parameters and the physical parameters, the Eqs.~\eqref{gauge boson fpeak0} and~\eqref{gauge boson Omega0} for the dark photon couplings, as well as Eqs.~\eqref{Nieh-Yan fc} and~\eqref{Nieh-Yan Omegac} for the Nieh-Yan coupling, 
which enables us to constrain the physical parameters via measurements of the spectral shape.

We introduce the dimensionless normalized model parameters by combining the axion mass $ m $, the coupling constant $ \alpha_{X/T} $, and the decay constant $ f_{X/T} $ as follows:
\begin{align}\label{mtilde}
&\tilde{m}_{X/T} = \frac{m}{\text{eV}}  
\left (  \frac{\alpha_{X/T}}{M_{P}^{2}  }  \right ) ^{4/3}, \\
&\tilde{f}_{X} = \frac{\fx }{M_{P} }  
\left ( \frac{\alpha_{X}}{M_{P}^{2} }  \right ) ^{-\frac{1}{3} },~~\tilde{f}_{T} = \frac{\fh }{M_{P} } , 
\end{align}
and substitute them into the spectral form~\eqref{eq:temp_AA} and~\eqref{BPL template}. 
From the specific physical parameters of the dark photon coupling and the Nieh-Yan coupling models, as described in Sec.~\ref{subsec:fit_temp_DP} and~\ref{subsec:fit_temp_NY} respectively, we obtain the values of these dimensionless normalized model parameters, which are explicitly listed in Table~\ref{tab:fundamentalparameter_values}. 
We use the Fisher matrix to estimate the normalized model parameters.

\begin{figure*}[htbp]
  \centering
  \captionsetup{justification=raggedright,singlelinecheck=false}
  \begin{subfigure}[b]{0.48\textwidth}
    \centering
    \includegraphics[width=\textwidth]{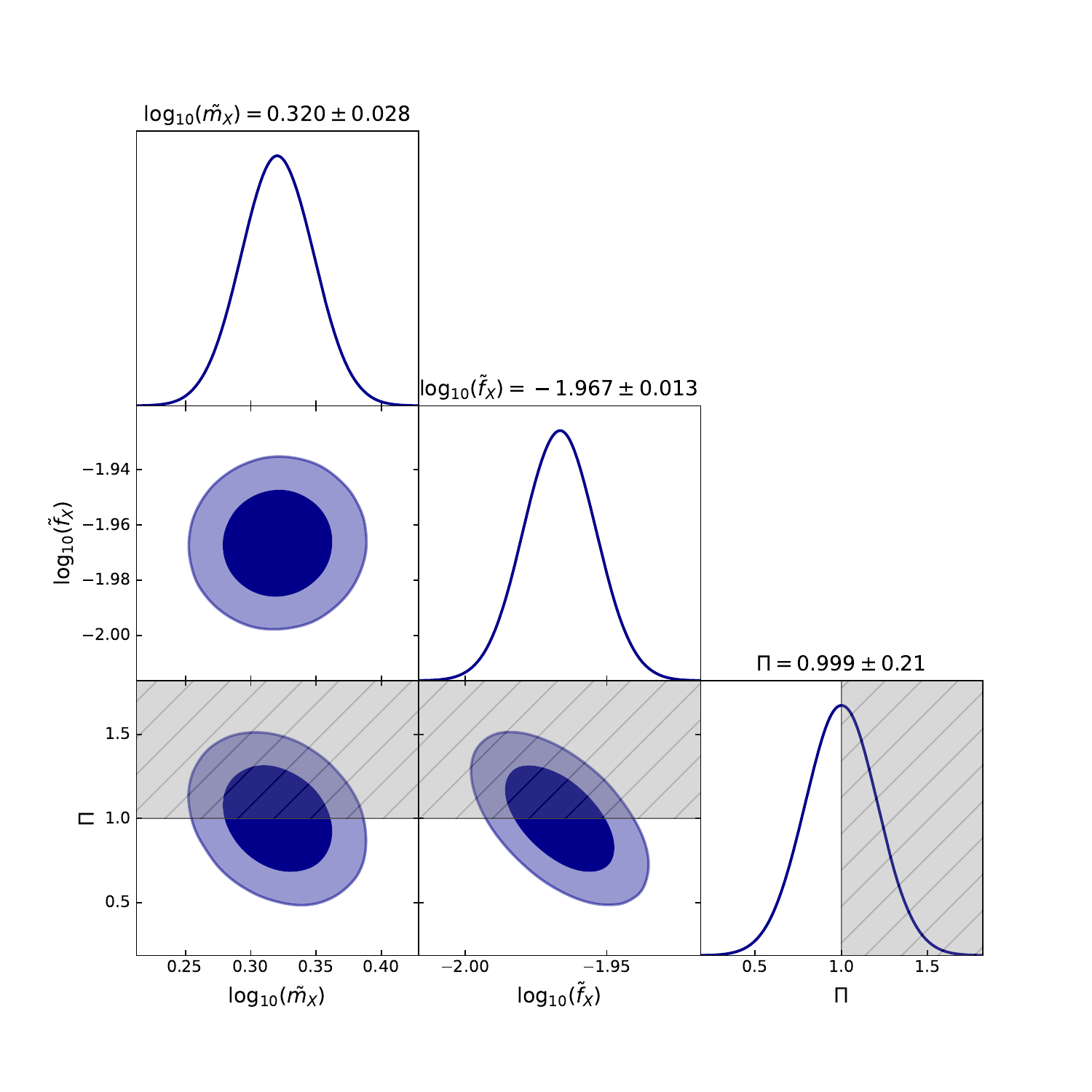}
    \caption{Normalized parameters in  dark photon coupling model.}
    \label{fig:corner_plots_AudibleAxio_Fundamentalparameters} 
  \end{subfigure}
  \hfill
\begin{subfigure}[b]{0.48\textwidth}
    \centering
    \includegraphics[width=\textwidth]{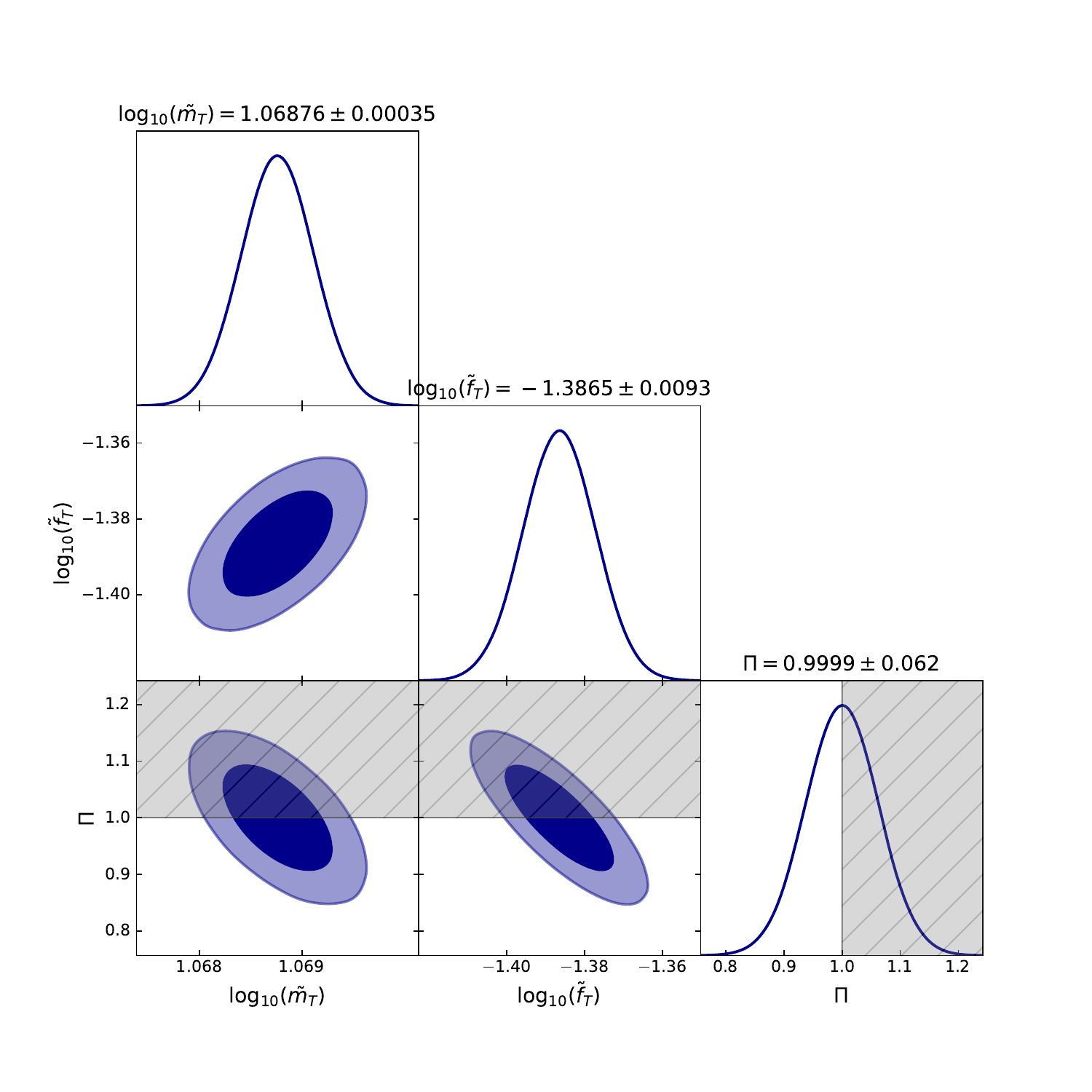}
    \caption{Normalized parameters in Nieh-Yan coupling model.}
    \label{Corner plots BPL_Fundamentalparameters} \end{subfigure}
  \caption{Corner plots showing SGWB normalized model parameters estimates from the Fisher matrix for (a) the dark photon coupling model and (b) the Nieh-Yan coupling model in Table~\ref{tab:fundamentalparameter_values}. 
  At the top of each column, the corresponding parameters' 1$\sigma$ uncertainties are presented. 
  The gray shaded areas correspond to regions of the parameter space with $\Pi > 1$ which is theoretically unacceptable.}
  \label{fig:combined_corner_plots}
\end{figure*}

We compute the Fisher matrix of the normalized model parameters $\tilde{m}$, $\tilde{f}_{X}$, and $\tilde{f}_{T}$ to obtain its covariance matrix. The correlation plots of this matrix are shown in Figs.~\ref{fig:corner_plots_AudibleAxio_Fundamentalparameters} and~\ref{Corner plots BPL_Fundamentalparameters}. We can constrain the values of the physical parameters by the measurement of these normalized model parameters.

For dark photon coupling model as given in Eq.~\eqref{eq:temp_AA}, the results are shown in Fig.~\ref{fig:corner_plots_AudibleAxio_Fundamentalparameters}. At the $1\sigma$ confidence level, the relative errors are less than $6.7\%$ for the normalized model parameters. It can be observed that the measurements of $\tilde{m}_{X}$ and $\tilde{f}_{X}$ are approximately independent, whereas $\Pi$ exhibits a certain degree of negative correlation with both $\tilde{m}_{X}$ and $\tilde{f}_{X}$. The parameter $\Pi$ is estimated to be $0.9999$ with a relative uncertainty of $21.0\%$ at the $1\sigma$ confidence level. To improve the measurement precision of $\Pi$, we can enhance the sensitivity of individual detectors and the detector network, as well as achieve higher SNR.

For the Nieh-Yan coupling model, the result for the broken power-law template in Eq.~\eqref{BPL template} is shown in Fig.~\ref{Corner plots BPL_Fundamentalparameters}. At the $1\sigma$ confidence level, the relative errors are less than $2.2\%$ for the normalized model parameters. It can be observed that the measurements of the parameters $\tilde{m}_{T}$, $\tilde{f}_{T}$, and $\Pi$ are correlated. Moreover, due to the strong signal strength, the measurement precision of $\Pi$ has been improved compared to the results for the dark photon coupling model, 
with a relative error of $6.2\%$ at the $1\sigma$ confidence level. Meanwhile, we can see that the measurement precision of $\tilde{m}_{T}$ is relatively high. This is because the broken power-law template has a narrow peak and strong amplitude at the peak frequency, measuring the peak frequency more sensitive. According to the relationships~\eqref{Nieh-Yan fc} and~\eqref{mtilde}, it is clear that $\tilde{m}$ corresponds to the measurement of the peak frequency. 

\begin{table}[H]
  \centering
  \begin{tabular}{|c |c| c| c|} %
\hline
   ${\cal{ O}}$~  &~~ $\tilde{m}_{{\cal O}} $ &~~~ $\tilde{f}_{{\cal O}}$ &~~~ $\Pi$ \\ \hline
    $X$(Dark Photon) &~~~ $2.092 $ &~~~ $ 0.0108 $ &~~~ $ 0.9999 $  \\ \hline
   $T$(Nieh-Yan) &~~~ $ 11.72 $ &~~~ $ 0.0411 $ &~~~ $ 0.9999 $  \\ 
  \hline
  \end{tabular}
\caption{Values of the normalized model parameters for dark photon coupling and Nieh-Yan coupling models. The subscript ${\cal O}$ in $\tilde{m}_{{\cal O}}$ and $\tilde{f}_{{\cal O}}$ denotes the model index: ${\cal O} = X$ for dark photon coupling, ${\cal O} = T$ for Nieh-Yan coupling.}
  \label{tab:fundamentalparameter_values}
\end{table}

\section{Conclusion}
\label{conclusion}
The single GW detectors face challenges in detecting the chirality of GWs due to their planar design. However, with the network of space-based detectors such as LISA and Taiji through cross-correlation techniques, we can compute chirality-dependent response functions and extract the net circular polarization of an isotropic SGWB. The detection of parity violation through chiral GWs is crucial for understanding the early Universe and for distinguishing a cosmological GW background from an astrophysical one~\cite{Orlando:2020oko}. 
In this work, we present the response functions for Stokes parameters $I$ and $V$, as well as the total intensity sensitivity curve for GWBs originating from audible axions, using the LISA-Taiji network. In addition to the LISA-Taiji network, other space-based GW detector networks such as LISA-TianQin have also been proposed and studied extensively~\cite{Gong:2021gvw, Li:2023szq, Wu:2023bwd, Luo:2025ewp}. 

We use the Fisher information matrix to estimate both spectral parameters and normalized model parameters of axion-induced chiral GW spectra through the LISA-Taiji network, focusing on axion-dark photon and axion-Nieh-Yan couplings with physical parameters selected to yield strong GWs in the mHz range. Our results demonstrate that the network estimates the spectral shape parameters and normalized model parameters for both coupling models. For the spectral shape parameters in the dark photon coupling model, we obtain relative errors of $62.0\%$ at the $1\sigma$ confidence level, while for the Nieh-Yan coupling model, the relative errors are less than $ 31.8\%$ at the same confidence level. Regarding circular polarization parameter $\Pi$, its relative error is less than $23.0\%$ ($1\sigma$) for the dark photon coupling model, and it is reduced to $6.7\%$ ($1\sigma$) for the stronger signals from the Nieh-Yan coupling model.
Compared to flat GW spectra in~\cite{Orlando:2020oko}, we have computed the parameter uncertainties for frequency-dependent spectra of the axion-induced chiral GWB, demonstrating the LISA-Taiji network's capability to effectively constrain both GW spectral parameters and normalized model parameters. 


\vspace{-2pt}

\section*{Acknowledgement}
This work is supported by the National Key Research and Development Program of China (No. 2023YFC2206200, No.2021YFC2201901) and
the National Natural Science Foundation of China (No.12375059, No.1240507). 

\vspace{-2pt}
\appendix

\section{Response functions of GWs}
\label{Appa}

In this work, we employ natural units and adopt the Lorentz transverse-traceless gauge. We establish a coordinate system $\left \{ \hat{e}_x, \hat{e}_y, \hat{e}_z \right \}$ at rest relative to an isotropic SGWB. 
\subsection{Polarization tensor basises}
For an incoming plane GW with a single wave vector $\veck$, we define an orthogonal basis
\begin{equation}
    \hat{u}(\hat{k})=\frac{\hat{k} \times \hat{e}_z}{|\hat{k} \times \hat{e}_z|}, \quad \hat{v}(\hat{k})=\hat{k} \times \hat{u},
\end{equation}
where $\hat{k}$ denotes the unit vector in the direction of wave-vector $\veck$, and its magnitude is given by $k = |\veck|$. 
Using the above equation, we define the so-called ``plus'' ($+$) and ``cross'' ($\times$) polarization tensors as
\begin{equation}
\begin{aligned}
e_{ab}^{+}(\hat{k}) = \frac{\hat{u}_a \hat{u}_b - \hat{v}_a \hat{v}_b}{\sqrt{2}} , \quad
e_{ab}^{\times}(\hat{k}) = \frac{\hat{u}_a \hat{v}_b + \hat{v}_a \hat{u}_b}{\sqrt{2}} .
\end{aligned}
\end{equation}

It is more convenient to introduce the circular polarization basis tensors $e_{ab}^R$ and $e_{ab}^L$ when searching for evidence of circular polarization in the background. 
Then the relationships between the left- and right-handed polarization tensors and the ``plus'' $(+)$ and ``cross'' $(\times)$ polarization basis are
\begin{equation}
    e_{a b}^R(\hat{k})=\frac{e_{a b}^{+}+i e_{a b}^{\times}}{\sqrt{2}}, \quad e_{a b}^L(\hat{k})=\frac{e_{a b}^{+}-i e_{a b}^{\times}}{\sqrt{2}}.
\end{equation}

The superposition of GWs arriving at position $\vec{x}$ at time $t$ can be represented as an incident plane wave
\begin{equation}
h_{ab}(\vec{x}, t) = \int_{-\infty}^{+\infty} \dd f \int_{\Omega} \dd \Omega_{\hat{k}} \, e^{2\pi i f(t - \hat{k} \cdot \vec{x})} \sum_{P} \tilde{h}_{P}(f, \hat{k}) e^{P}_{ab}(\hat{k}),
\end{equation}
where the index $P$ labels either the plus and cross polarizations ($+$/$\times$) or the left- and right-handed polarizations (L/R). Here, $f = k\,c$ denotes the frequency of each plane wave, $\dd\Omega_{\hat{k}}$ represents the infinitesimal solid angle corresponding to the wave vector $\veck$, and $\tilde{h}_{P}(f, \hat{k}) \equiv f^2 \tilde{h}_{P}(\veck)$. Finally, the gravitational wave can be expressed in terms of $\veck$ as
\begin{equation}
\begin{aligned}
h_{ab}(\vec{x}, t) = & \int \mathrm{d}^3 k \, e^{-2 \pi i \veck \cdot \vec{x}} \sum_P \left[ e^{2 \pi i k t} \tilde{h}_P(\veck) e_{ab}^P(\hat{k}) \right. \\
& \left. + e^{-2 \pi i k t} \tilde{h}_P^*(-\veck) e_{ab}^{P *}(-\hat{k}) \right].
\end{aligned}
\end{equation}

\subsection{Quadratic response functions}
\label{appendix:Response_functions}

In actual measurements, space-based detectors measure differential Doppler frequency shifts rather than direct time shifts. These shifts are defined as $\Delta F_{12}(t) \equiv \Delta \nu_{12}(t)/\nu = -d \Delta T_{12}(t)/dt$. 
We use $L$ to denote the detector arm length.
The most straightforward interferometric measurement at a vertex performed by a space-based detector is
\begin{equation}
\begin{aligned}
\Delta F_{1(23)}(t) &= \Delta F_{21}(t - L) + \Delta F_{12}(t) \\
&\quad - \left[\Delta F_{31}(t - L) + \Delta F_{13}(t)\right].
\end{aligned}
\end{equation}

In order to suppress noise induced by laser phase variations and other factors, we implement TDI techniques. 
Consider two test masses labeled $i$ and $j$, and let $\hat{l}_{ij} = \left ( {\bf x}_{j} - {\bf x}_{i} \right ) /\left | {\bf x}_{j} - {\bf x}_{i} \right | $ denote the unit vector pointing from mass $i$ to mass $j$ among the three detector spacecraft. 
The TDI1.5 variable is obtained through cyclic permutation of the TDI variables Y and Z
\begin{equation}
\begin{aligned}
&\Delta F^{1.5}_{1(23)}(t) = \Delta F_{1(23)}(t - 2L) + \Delta F_{1(32)}(t) \\
=& - \int \dd^3 k \, e^{-2\pi i \veck \cdot \vec{x}_{1}} (2\pi i k L) \times  \\
& \sum_{\lambda} \left[ 
    e^{2\pi i k (t - L)} W(k L) R^{\lambda}_{1}(\veck,\hat{l}_{12},\hat{l}_{13}) \tilde{h}_{\lambda}(k) \right. \\
&\quad \left. - e^{-2\pi i k (t - L)} W^{\ast}(k L) R^{\lambda^{\ast}}_{1}(-\veck,\hat{l}_{12},\hat{l}_{13}) \tilde{h}^{\ast}_{\lambda}(-k) \right] .
\end{aligned}
\end{equation}
where $\lambda = L$ or $R$ denotes left- and right-handed polarizations, and $W(kL) \equiv e^{-4\pi ikL} - 1$.
The function $R^{\lambda}_{i}$ is defined as
\begin{equation}
\begin{aligned}
R_i^\lambda (\veck, \hat{l}_{i j}, \hat{l}_{i k} )& 
\equiv   \frac{\hat{l}_{i j}^a \hat{l}_{i j}^b}{2} e_{a b}^\lambda(\hat{k}) \mathcal{T}(\veck, \hat{l}_{i j}) 
-\frac{\hat{l}_{i k}^a \hat{l}_{i k}^b}{2} e_{a b}^\lambda(\hat{k}) \mathcal{T}(\veck, \hat{l}_{i k}),
\end{aligned}
\end{equation}
where the detector transfer function $\mathcal{T}(\veck, \hat{l}_{i j})$ is given by
\begin{equation}
\begin{aligned}
\mathcal{T}(\veck, \hat{l}_{i j})& \equiv  \mathrm{e}^{\pi i k L\left(1-\hat{k} \cdot \hat{l}_{i j}\right)} \operatorname{sinc}\left[\pi k L\left(1+\hat{k} \cdot \hat{l}_{i j}\right)\right] \\
+&\,  \mathrm{e}^{-\pi i k L\left(1+\hat{k} \cdot \hat{l}_{i j}\right)} \operatorname{sinc}\left[\pi k L\left(1-\hat{k} \cdot \hat{l}_{i j}\right)\right].
\end{aligned}
\end{equation}

For simplicity, we represent the detector output using the notation $s_i(t) \equiv \Delta F^{1.5}_{i(jk)}(t)$. The information is contained within the two-point correlation functions of the data streams. Below, without assuming identical detectors, we present the general formulation. The two-point cross-correlation is expressed as
\begin{equation}
\begin{aligned}
&\left \langle s_{i}(t) s_{j}(t) \right \rangle = \int \dd k \, (2 \pi k L_{i}) (2 \pi k L_{j}) \sum_{\lambda} P_{\lambda}(k) \times   \\
&~ \left[ e^{-2 \pi i k(L_{i} - L_{j})} W(k L_{i}) W^{\ast}(k L_{j}) 
\tilde{\Gamma}_{ij}^{\lambda}(k) + \text{h.c.} \right],
\end{aligned}
\end{equation}
where the cross-correlation function
\begin{equation}
\begin{aligned}\label{GeoResponseFunction}
\tilde{\Gamma}_{i j}^\lambda(k) \equiv & \, \frac{1}{4 \pi} \int \mathrm{d}^2 \hat{k} \, \mathrm{e}^{-2 \pi i \veck \cdot\left(\vec{x}_i-\vec{x}_j\right)} \\
& \times \, R_i^\lambda\left(\veck, \hat{l}_{i k}, \hat{l}_{i l}\right) R_j^{\lambda^*}\left(\veck, \hat{l}_{j m}, \hat{l}_{j n}\right) .
\end{aligned}
\end{equation}
For notational simplicity, the quadratic response function
\begin{equation}
\begin{aligned}\label{ResponseFunction}
\Gamma_{ij}^{\lambda}(k) \equiv &\, (2\pi k L_i)(2\pi k L_j) W(kL_i)W^*(kL_j) 
\tilde{\Gamma}_{ij}^{\lambda}(k) + \text{h.c.}
\;,
\end{aligned}
\end{equation}
we obtain the compact form
\begin{equation}
\langle s_{i}(t) s_{j}(t) \rangle = \int \dd k \left[ \Gamma_{ij}^{L}(k) P_{L}(k) + \Gamma_{ij}^{R}(k) P_{R}(k) \right] .
\end{equation}
When $i$ and $j$ are both LISA (or Taiji) channels, the response function is given by
\begin{equation}
\begin{aligned}\label{SelfResponseFunction}
\Gamma_{ij}(k)
&= \Gamma^L_{ij}(k) + \Gamma^R_{ij}(k) \\
&= 16(2\pi k L)^2 \sin^2(2\pi k L) \tilde{\Gamma}_{ij}(k),
\end{aligned}
\end{equation}
where $\tilde{\Gamma}_{i j}(k) = \tilde{\Gamma}^L_{ij}(k) = \tilde{\Gamma}^R_{ij}(k)$~\cite{Flauger:2020qyi}. 
The response functions $\Gamma_{ij}$ for the A and E channels of LISA and Taiji are shown in Fig.~\ref{fig:Response_functions}. 

Using the methods in~\cite{Robson:2018ifk,Orlando:2020oko}, we have calculated the response functions and total intensity sensitivity curves associated with the self-correlation of LISA and Taiji. The corresponding results are shown in Fig.~\ref{fig:combined_response_LISA}. 
Additionally, we have obtained the response functions for the $I$ and $V$ components of all cross-correlation TDI channels between LISA and Taiji, as shown in Fig.~\ref{fig:The geometric factor of the response function}. For further details on the derivation, we refer to~\cite{Orlando:2020oko,Flauger:2020qyi}.

\subsection{The AET basises}

In space-based gravitational wave detection, the fundamental TDI channels of a Michelson interferometer include X, Y, and Z. 
We transform the X, Y, and Z channels into the noise-independent channels A, E, and T, which are related as follows~\cite{Orlando:2020oko}
\begin{align}
\tilde{d}_A &= \frac{1}{\sqrt{2}} (\tilde{d}_Z - \tilde{d}_X), \\
\tilde{d}_E &= \frac{1}{\sqrt{6}} (\tilde{d}_X - 2\tilde{d}_Y + \tilde{d}_Z), \\
\tilde{d}_T &= \frac{1}{\sqrt{3}} (\tilde{d}_X + \tilde{d}_Y + \tilde{d}_Z).
\end{align}
The noise independence of the A, E, and T channels is valid under the assumptions of identical noise in each laser link and equal arm lengths.

For convenience, we summarize the key symbols used in this paper in Table~\ref{table:Symbol}.
\begin{table}[H]
\centering
\renewcommand{\arraystretch}{1.5}  
\setlength{\tabcolsep}{4pt} 
\begin{tabularx}{\columnwidth}{|c|X|}
\hline
Symbols & Descriptions \\
\hline
$S_{ij}(f)$ & One-sided signal PSD \\
\hline
$N_{i}(f)$ & One-sided noise PSD \\
\hline
$h_{ab}(\vec{x}, t)$ & Gravitational wave strain tensor \\
\hline
$\Delta F_{ij}(t)$ & Doppler frequency shifts \\
\hline
$\tilde{\Gamma}_{i j}^\lambda(k)$ & Cross-correlation function\\
\hline
$\Gamma_{i j}^\lambda(k)$ &  Quadratic response function \\
\hline
\end{tabularx}
\caption{Summary of the symbols and descriptions.}
\label{table:Symbol}
\end{table}

\bibliography{references.bib}

\begin{thebibliography}{99}%
\makeatletter
\providecommand \@ifxundefined [1]{%
 \@ifx{#1\undefined}
}%
\providecommand \@ifnum [1]{%
 \ifnum #1\expandafter \@firstoftwo
 \else \expandafter \@secondoftwo
 \fi
}%
\providecommand \@ifx [1]{%
 \ifx #1\expandafter \@firstoftwo
 \else \expandafter \@secondoftwo
 \fi
}%
\providecommand \natexlab [1]{#1}%
\providecommand \enquote  [1]{``#1''}%
\providecommand \bibnamefont  [1]{#1}%
\providecommand \bibfnamefont [1]{#1}%
\providecommand \citenamefont [1]{#1}%
\providecommand \href@noop [0]{\@secondoftwo}%
\providecommand \href [0]{\begingroup \@sanitize@url \@href}%
\providecommand \@href[1]{\@@startlink{#1}\@@href}%
\providecommand \@@href[1]{\endgroup#1\@@endlink}%
\providecommand \@sanitize@url [0]{\catcode `\\12\catcode `\$12\catcode
  `\&12\catcode `\#12\catcode `\^12\catcode `\_12\catcode `\%12\relax}%
\providecommand \@@startlink[1]{}%
\providecommand \@@endlink[0]{}%
\providecommand \url  [0]{\begingroup\@sanitize@url \@url }%
\providecommand \@url [1]{\endgroup\@href {#1}{\urlprefix }}%
\providecommand \urlprefix  [0]{URL }%
\providecommand \Eprint [0]{\href }%
\providecommand \doibase [0]{https://doi.org/}%
\providecommand \selectlanguage [0]{\@gobble}%
\providecommand \bibinfo  [0]{\@secondoftwo}%
\providecommand \bibfield  [0]{\@secondoftwo}%
\providecommand \translation [1]{[#1]}%
\providecommand \BibitemOpen [0]{}%
\providecommand \bibitemStop [0]{}%
\providecommand \bibitemNoStop [0]{.\EOS\space}%
\providecommand \EOS [0]{\spacefactor3000\relax}%
\providecommand \BibitemShut  [1]{\csname bibitem#1\endcsname}%
\let\auto@bib@innerbib\@empty
\bibitem [{\citenamefont {Abbott}\ \emph {et~al.}(2016)\citenamefont {Abbott}
  \emph {et~al.}}]{LIGOScientific:2016aoc}%
  \BibitemOpen
  \bibfield  {author} {\bibinfo {author} {\bibfnamefont {B.~P.}\ \bibnamefont
  {Abbott}} \emph {et~al.} (\bibinfo {collaboration} {LIGO Scientific,
  Virgo}),\ }\bibfield  {title} {\bibinfo {title} {{Observation of
  Gravitational Waves from a Binary Black Hole Merger}},\ }\href
  {https://doi.org/10.1103/PhysRevLett.116.061102} {\bibfield  {journal}
  {\bibinfo  {journal} {Phys. Rev. Lett.}\ }\textbf {\bibinfo {volume} {116}},\
  \bibinfo {pages} {061102} (\bibinfo {year} {2016})},\ \Eprint
  {https://arxiv.org/abs/1602.03837} {arXiv:1602.03837 [gr-qc]} \BibitemShut
  {NoStop}%
\bibitem [{\citenamefont {Abramovici}\ \emph {et~al.}(1992)\citenamefont
  {Abramovici} \emph {et~al.}}]{Abramovici:1992ah}%
  \BibitemOpen
  \bibfield  {author} {\bibinfo {author} {\bibfnamefont {A.}~\bibnamefont
  {Abramovici}} \emph {et~al.},\ }\bibfield  {title} {\bibinfo {title} {{LIGO:
  The Laser interferometer gravitational wave observatory}},\ }\href
  {https://doi.org/10.1126/science.256.5055.325} {\bibfield  {journal}
  {\bibinfo  {journal} {Science}\ }\textbf {\bibinfo {volume} {256}},\ \bibinfo
  {pages} {325} (\bibinfo {year} {1992})}\BibitemShut {NoStop}%
\bibitem [{\citenamefont {Allen}(1996)}]{Allen:1996vm}%
  \BibitemOpen
  \bibfield  {author} {\bibinfo {author} {\bibfnamefont {B.}~\bibnamefont
  {Allen}},\ }\bibfield  {title} {\bibinfo {title} {{The Stochastic gravity
  wave background: Sources and detection}},\ }in\ \href@noop {} {\emph
  {\bibinfo {booktitle} {{Les Houches School of Physics: Astrophysical Sources
  of Gravitational Radiation}}}}\ (\bibinfo {year} {1996})\ pp.\ \bibinfo
  {pages} {373--417},\ \Eprint {https://arxiv.org/abs/gr-qc/9604033}
  {arXiv:gr-qc/9604033} \BibitemShut {NoStop}%
\bibitem [{\citenamefont {Maggiore}(2000)}]{Maggiore:1999vm}%
  \BibitemOpen
  \bibfield  {author} {\bibinfo {author} {\bibfnamefont {M.}~\bibnamefont
  {Maggiore}},\ }\bibfield  {title} {\bibinfo {title} {{Gravitational wave
  experiments and early universe cosmology}},\ }\href
  {https://doi.org/10.1016/S0370-1573(99)00102-7} {\bibfield  {journal}
  {\bibinfo  {journal} {Phys. Rept.}\ }\textbf {\bibinfo {volume} {331}},\
  \bibinfo {pages} {283} (\bibinfo {year} {2000})},\ \Eprint
  {https://arxiv.org/abs/gr-qc/9909001} {arXiv:gr-qc/9909001} \BibitemShut
  {NoStop}%
\bibitem [{\citenamefont {Kuroyanagi}\ \emph {et~al.}(2018)\citenamefont
  {Kuroyanagi}, \citenamefont {Chiba},\ and\ \citenamefont
  {Takahashi}}]{Kuroyanagi:2018csn}%
  \BibitemOpen
  \bibfield  {author} {\bibinfo {author} {\bibfnamefont {S.}~\bibnamefont
  {Kuroyanagi}}, \bibinfo {author} {\bibfnamefont {T.}~\bibnamefont {Chiba}},\
  and\ \bibinfo {author} {\bibfnamefont {T.}~\bibnamefont {Takahashi}},\
  }\bibfield  {title} {\bibinfo {title} {{Probing the Universe through the
  Stochastic Gravitational Wave Background}},\ }\href
  {https://doi.org/10.1088/1475-7516/2018/11/038} {\bibfield  {journal}
  {\bibinfo  {journal} {JCAP}\ }\textbf {\bibinfo {volume} {11}},\ \bibinfo
  {pages} {038}},\ \Eprint {https://arxiv.org/abs/1807.00786} {arXiv:1807.00786
  [astro-ph.CO]} \BibitemShut {NoStop}%
\bibitem [{\citenamefont {Auclair}\ \emph {et~al.}(2023)\citenamefont {Auclair}
  \emph {et~al.}}]{LISACosmologyWorkingGroup:2022jok}%
  \BibitemOpen
  \bibfield  {author} {\bibinfo {author} {\bibfnamefont {P.}~\bibnamefont
  {Auclair}} \emph {et~al.} (\bibinfo {collaboration} {LISA Cosmology Working
  Group}),\ }\bibfield  {title} {\bibinfo {title} {{Cosmology with the Laser
  Interferometer Space Antenna}},\ }\href
  {https://doi.org/10.1007/s41114-023-00045-2} {\bibfield  {journal} {\bibinfo
  {journal} {Living Rev. Rel.}\ }\textbf {\bibinfo {volume} {26}},\ \bibinfo
  {pages} {5} (\bibinfo {year} {2023})},\ \Eprint
  {https://arxiv.org/abs/2204.05434} {arXiv:2204.05434 [astro-ph.CO]}
  \BibitemShut {NoStop}%
\bibitem [{\citenamefont {Binetruy}\ \emph {et~al.}(2012)\citenamefont
  {Binetruy}, \citenamefont {Bohe}, \citenamefont {Caprini},\ and\
  \citenamefont {Dufaux}}]{Binetruy:2012ze}%
  \BibitemOpen
  \bibfield  {author} {\bibinfo {author} {\bibfnamefont {P.}~\bibnamefont
  {Binetruy}}, \bibinfo {author} {\bibfnamefont {A.}~\bibnamefont {Bohe}},
  \bibinfo {author} {\bibfnamefont {C.}~\bibnamefont {Caprini}},\ and\ \bibinfo
  {author} {\bibfnamefont {J.-F.}\ \bibnamefont {Dufaux}},\ }\bibfield  {title}
  {\bibinfo {title} {{Cosmological Backgrounds of Gravitational Waves and
  eLISA/NGO: Phase Transitions, Cosmic Strings and Other Sources}},\ }\href
  {https://doi.org/10.1088/1475-7516/2012/06/027} {\bibfield  {journal}
  {\bibinfo  {journal} {JCAP}\ }\textbf {\bibinfo {volume} {06}},\ \bibinfo
  {pages} {027}},\ \Eprint {https://arxiv.org/abs/1201.0983} {arXiv:1201.0983
  [gr-qc]} \BibitemShut {NoStop}%
\bibitem [{\citenamefont {Thrane}\ and\ \citenamefont
  {Romano}(2013)}]{Thrane:2013oya}%
  \BibitemOpen
  \bibfield  {author} {\bibinfo {author} {\bibfnamefont {E.}~\bibnamefont
  {Thrane}}\ and\ \bibinfo {author} {\bibfnamefont {J.~D.}\ \bibnamefont
  {Romano}},\ }\bibfield  {title} {\bibinfo {title} {{Sensitivity curves for
  searches for gravitational-wave backgrounds}},\ }\href
  {https://doi.org/10.1103/PhysRevD.88.124032} {\bibfield  {journal} {\bibinfo
  {journal} {Phys. Rev. D}\ }\textbf {\bibinfo {volume} {88}},\ \bibinfo
  {pages} {124032} (\bibinfo {year} {2013})},\ \Eprint
  {https://arxiv.org/abs/1310.5300} {arXiv:1310.5300 [astro-ph.IM]}
  \BibitemShut {NoStop}%
\bibitem [{\citenamefont {Romano}\ and\ \citenamefont
  {Cornish}(2017)}]{Romano:2016dpx}%
  \BibitemOpen
  \bibfield  {author} {\bibinfo {author} {\bibfnamefont {J.~D.}\ \bibnamefont
  {Romano}}\ and\ \bibinfo {author} {\bibfnamefont {N.~J.}\ \bibnamefont
  {Cornish}},\ }\bibfield  {title} {\bibinfo {title} {{Detection methods for
  stochastic gravitational-wave backgrounds: a unified treatment}},\ }\href
  {https://doi.org/10.1007/s41114-017-0004-1} {\bibfield  {journal} {\bibinfo
  {journal} {Living Rev. Rel.}\ }\textbf {\bibinfo {volume} {20}},\ \bibinfo
  {pages} {2} (\bibinfo {year} {2017})},\ \Eprint
  {https://arxiv.org/abs/1608.06889} {arXiv:1608.06889 [gr-qc]} \BibitemShut
  {NoStop}%
\bibitem [{\citenamefont {Caprini}\ and\ \citenamefont
  {Figueroa}(2018)}]{Caprini:2018mtu}%
  \BibitemOpen
  \bibfield  {author} {\bibinfo {author} {\bibfnamefont {C.}~\bibnamefont
  {Caprini}}\ and\ \bibinfo {author} {\bibfnamefont {D.~G.}\ \bibnamefont
  {Figueroa}},\ }\bibfield  {title} {\bibinfo {title} {{Cosmological
  Backgrounds of Gravitational Waves}},\ }\href
  {https://doi.org/10.1088/1361-6382/aac608} {\bibfield  {journal} {\bibinfo
  {journal} {Class. Quant. Grav.}\ }\textbf {\bibinfo {volume} {35}},\ \bibinfo
  {pages} {163001} (\bibinfo {year} {2018})},\ \Eprint
  {https://arxiv.org/abs/1801.04268} {arXiv:1801.04268 [astro-ph.CO]}
  \BibitemShut {NoStop}%
\bibitem [{\citenamefont {Boileau}\ \emph {et~al.}(2021)\citenamefont
  {Boileau}, \citenamefont {Christensen}, \citenamefont {Meyer},\ and\
  \citenamefont {Cornish}}]{Boileau:2020rpg}%
  \BibitemOpen
  \bibfield  {author} {\bibinfo {author} {\bibfnamefont {G.}~\bibnamefont
  {Boileau}}, \bibinfo {author} {\bibfnamefont {N.}~\bibnamefont
  {Christensen}}, \bibinfo {author} {\bibfnamefont {R.}~\bibnamefont {Meyer}},\
  and\ \bibinfo {author} {\bibfnamefont {N.~J.}\ \bibnamefont {Cornish}},\
  }\bibfield  {title} {\bibinfo {title} {{Spectral separation of the stochastic
  gravitational-wave background for LISA: Observing both cosmological and
  astrophysical backgrounds}},\ }\href
  {https://doi.org/10.1103/PhysRevD.103.103529} {\bibfield  {journal} {\bibinfo
   {journal} {Phys. Rev. D}\ }\textbf {\bibinfo {volume} {103}},\ \bibinfo
  {pages} {103529} (\bibinfo {year} {2021})},\ \Eprint
  {https://arxiv.org/abs/2011.05055} {arXiv:2011.05055 [gr-qc]} \BibitemShut
  {NoStop}%
\bibitem [{\citenamefont {Flauger}\ \emph {et~al.}(2021)\citenamefont
  {Flauger}, \citenamefont {Karnesis}, \citenamefont {Nardini}, \citenamefont
  {Pieroni}, \citenamefont {Ricciardone},\ and\ \citenamefont
  {Torrado}}]{Flauger:2020qyi}%
  \BibitemOpen
  \bibfield  {author} {\bibinfo {author} {\bibfnamefont {R.}~\bibnamefont
  {Flauger}}, \bibinfo {author} {\bibfnamefont {N.}~\bibnamefont {Karnesis}},
  \bibinfo {author} {\bibfnamefont {G.}~\bibnamefont {Nardini}}, \bibinfo
  {author} {\bibfnamefont {M.}~\bibnamefont {Pieroni}}, \bibinfo {author}
  {\bibfnamefont {A.}~\bibnamefont {Ricciardone}},\ and\ \bibinfo {author}
  {\bibfnamefont {J.}~\bibnamefont {Torrado}},\ }\bibfield  {title} {\bibinfo
  {title} {{Improved reconstruction of a stochastic gravitational wave
  background with LISA}},\ }\href
  {https://doi.org/10.1088/1475-7516/2021/01/059} {\bibfield  {journal}
  {\bibinfo  {journal} {JCAP}\ }\textbf {\bibinfo {volume} {01}},\ \bibinfo
  {pages} {059}},\ \Eprint {https://arxiv.org/abs/2009.11845} {arXiv:2009.11845
  [astro-ph.CO]} \BibitemShut {NoStop}%
\bibitem [{\citenamefont {van Remortel}\ \emph {et~al.}(2023)\citenamefont {van
  Remortel}, \citenamefont {Janssens},\ and\ \citenamefont
  {Turbang}}]{vanRemortel:2022fkb}%
  \BibitemOpen
  \bibfield  {author} {\bibinfo {author} {\bibfnamefont {N.}~\bibnamefont {van
  Remortel}}, \bibinfo {author} {\bibfnamefont {K.}~\bibnamefont {Janssens}},\
  and\ \bibinfo {author} {\bibfnamefont {K.}~\bibnamefont {Turbang}},\
  }\bibfield  {title} {\bibinfo {title} {{Stochastic gravitational wave
  background: Methods and implications}},\ }\href
  {https://doi.org/10.1016/j.ppnp.2022.104003} {\bibfield  {journal} {\bibinfo
  {journal} {Prog. Part. Nucl. Phys.}\ }\textbf {\bibinfo {volume} {128}},\
  \bibinfo {pages} {104003} (\bibinfo {year} {2023})},\ \Eprint
  {https://arxiv.org/abs/2210.00761} {arXiv:2210.00761 [gr-qc]} \BibitemShut
  {NoStop}%
\bibitem [{\citenamefont {Peccei}\ and\ \citenamefont
  {Quinn}(1977{\natexlab{a}})}]{Peccei1977hh}%
  \BibitemOpen
  \bibfield  {author} {\bibinfo {author} {\bibfnamefont {R.~D.}\ \bibnamefont
  {Peccei}}\ and\ \bibinfo {author} {\bibfnamefont {H.~R.}\ \bibnamefont
  {Quinn}},\ }\bibfield  {title} {\bibinfo {title} {{CP Conservation in the
  Presence of Instantons}},\ }\href
  {https://doi.org/10.1103/PhysRevLett.38.1440} {\bibfield  {journal} {\bibinfo
   {journal} {Phys. Rev. Lett.}\ }\textbf {\bibinfo {volume} {38}},\ \bibinfo
  {pages} {1440} (\bibinfo {year} {1977}{\natexlab{a}})}\BibitemShut {NoStop}%
\bibitem [{\citenamefont {Peccei}\ and\ \citenamefont
  {Quinn}(1977{\natexlab{b}})}]{Peccei:1977ur}%
  \BibitemOpen
  \bibfield  {author} {\bibinfo {author} {\bibfnamefont {R.~D.}\ \bibnamefont
  {Peccei}}\ and\ \bibinfo {author} {\bibfnamefont {H.~R.}\ \bibnamefont
  {Quinn}},\ }\bibfield  {title} {\bibinfo {title} {{Constraints Imposed by CP
  Conservation in the Presence of Instantons}},\ }\href
  {https://doi.org/10.1103/PhysRevD.16.1791} {\bibfield  {journal} {\bibinfo
  {journal} {Phys. Rev. D}\ }\textbf {\bibinfo {volume} {16}},\ \bibinfo
  {pages} {1791} (\bibinfo {year} {1977}{\natexlab{b}})}\BibitemShut {NoStop}%
\bibitem [{\citenamefont {Weinberg}(1978)}]{Weinberg:1977ma}%
  \BibitemOpen
  \bibfield  {author} {\bibinfo {author} {\bibfnamefont {S.}~\bibnamefont
  {Weinberg}},\ }\bibfield  {title} {\bibinfo {title} {{A New Light Boson?}},\
  }\href {https://doi.org/10.1103/PhysRevLett.40.223} {\bibfield  {journal}
  {\bibinfo  {journal} {Phys. Rev. Lett.}\ }\textbf {\bibinfo {volume} {40}},\
  \bibinfo {pages} {223} (\bibinfo {year} {1978})}\BibitemShut {NoStop}%
\bibitem [{\citenamefont {Wilczek}(1978)}]{Wilczek:1977pj}%
  \BibitemOpen
  \bibfield  {author} {\bibinfo {author} {\bibfnamefont {F.}~\bibnamefont
  {Wilczek}},\ }\bibfield  {title} {\bibinfo {title} {{Problem of Strong $P$
  and $T$ Invariance in the Presence of Instantons}},\ }\href
  {https://doi.org/10.1103/PhysRevLett.40.279} {\bibfield  {journal} {\bibinfo
  {journal} {Phys. Rev. Lett.}\ }\textbf {\bibinfo {volume} {40}},\ \bibinfo
  {pages} {279} (\bibinfo {year} {1978})}\BibitemShut {NoStop}%
\bibitem [{\citenamefont {Du}\ \emph {et~al.}(2018)\citenamefont {Du} \emph
  {et~al.}}]{ADMX:2018gho}%
  \BibitemOpen
  \bibfield  {author} {\bibinfo {author} {\bibfnamefont {N.}~\bibnamefont {Du}}
  \emph {et~al.} (\bibinfo {collaboration} {ADMX}),\ }\bibfield  {title}
  {\bibinfo {title} {{A Search for Invisible Axion Dark Matter with the Axion
  Dark Matter Experiment}},\ }\href
  {https://doi.org/10.1103/PhysRevLett.120.151301} {\bibfield  {journal}
  {\bibinfo  {journal} {Phys. Rev. Lett.}\ }\textbf {\bibinfo {volume} {120}},\
  \bibinfo {pages} {151301} (\bibinfo {year} {2018})},\ \Eprint
  {https://arxiv.org/abs/1804.05750} {arXiv:1804.05750 [hep-ex]} \BibitemShut
  {NoStop}%
\bibitem [{\citenamefont {Di~Luzio}\ \emph {et~al.}(2020)\citenamefont
  {Di~Luzio}, \citenamefont {Giannotti}, \citenamefont {Nardi},\ and\
  \citenamefont {Visinelli}}]{DiLuzio:2020wdo}%
  \BibitemOpen
  \bibfield  {author} {\bibinfo {author} {\bibfnamefont {L.}~\bibnamefont
  {Di~Luzio}}, \bibinfo {author} {\bibfnamefont {M.}~\bibnamefont {Giannotti}},
  \bibinfo {author} {\bibfnamefont {E.}~\bibnamefont {Nardi}},\ and\ \bibinfo
  {author} {\bibfnamefont {L.}~\bibnamefont {Visinelli}},\ }\bibfield  {title}
  {\bibinfo {title} {{The landscape of QCD axion models}},\ }\href
  {https://doi.org/10.1016/j.physrep.2020.06.002} {\bibfield  {journal}
  {\bibinfo  {journal} {Phys. Rept.}\ }\textbf {\bibinfo {volume} {870}},\
  \bibinfo {pages} {1} (\bibinfo {year} {2020})},\ \Eprint
  {https://arxiv.org/abs/2003.01100} {arXiv:2003.01100 [hep-ph]} \BibitemShut
  {NoStop}%
\bibitem [{\citenamefont {Abbott}\ and\ \citenamefont
  {Sikivie}(1983)}]{Abbott:1982af}%
  \BibitemOpen
  \bibfield  {author} {\bibinfo {author} {\bibfnamefont {L.~F.}\ \bibnamefont
  {Abbott}}\ and\ \bibinfo {author} {\bibfnamefont {P.}~\bibnamefont
  {Sikivie}},\ }\bibfield  {title} {\bibinfo {title} {{A Cosmological Bound on
  the Invisible Axion}},\ }\href {https://doi.org/10.1016/0370-2693(83)90638-X}
  {\bibfield  {journal} {\bibinfo  {journal} {Phys. Lett. B}\ }\textbf
  {\bibinfo {volume} {120}},\ \bibinfo {pages} {133} (\bibinfo {year}
  {1983})}\BibitemShut {NoStop}%
\bibitem [{\citenamefont {Ipser}\ and\ \citenamefont
  {Sikivie}(1983)}]{Ipser:1983mw}%
  \BibitemOpen
  \bibfield  {author} {\bibinfo {author} {\bibfnamefont {J.}~\bibnamefont
  {Ipser}}\ and\ \bibinfo {author} {\bibfnamefont {P.}~\bibnamefont
  {Sikivie}},\ }\bibfield  {title} {\bibinfo {title} {{Are Galactic Halos Made
  of Axions?}},\ }\href {https://doi.org/10.1103/PhysRevLett.50.925} {\bibfield
   {journal} {\bibinfo  {journal} {Phys. Rev. Lett.}\ }\textbf {\bibinfo
  {volume} {50}},\ \bibinfo {pages} {925} (\bibinfo {year} {1983})}\BibitemShut
  {NoStop}%
\bibitem [{\citenamefont {Marsh}(2016)}]{Marsh:2015xka}%
  \BibitemOpen
  \bibfield  {author} {\bibinfo {author} {\bibfnamefont {D.~J.~E.}\
  \bibnamefont {Marsh}},\ }\bibfield  {title} {\bibinfo {title} {{Axion
  Cosmology}},\ }\href {https://doi.org/10.1016/j.physrep.2016.06.005}
  {\bibfield  {journal} {\bibinfo  {journal} {Phys. Rept.}\ }\textbf {\bibinfo
  {volume} {643}},\ \bibinfo {pages} {1} (\bibinfo {year} {2016})},\ \Eprint
  {https://arxiv.org/abs/1510.07633} {arXiv:1510.07633 [astro-ph.CO]}
  \BibitemShut {NoStop}%
\bibitem [{\citenamefont {Preskill}\ \emph {et~al.}(1983)\citenamefont
  {Preskill}, \citenamefont {Wise},\ and\ \citenamefont
  {Wilczek}}]{Preskill:1982cy}%
  \BibitemOpen
  \bibfield  {author} {\bibinfo {author} {\bibfnamefont {J.}~\bibnamefont
  {Preskill}}, \bibinfo {author} {\bibfnamefont {M.~B.}\ \bibnamefont {Wise}},\
  and\ \bibinfo {author} {\bibfnamefont {F.}~\bibnamefont {Wilczek}},\
  }\bibfield  {title} {\bibinfo {title} {{Cosmology of the Invisible Axion}},\
  }\href {https://doi.org/10.1016/0370-2693(83)90637-8} {\bibfield  {journal}
  {\bibinfo  {journal} {Phys. Lett. B}\ }\textbf {\bibinfo {volume} {120}},\
  \bibinfo {pages} {127} (\bibinfo {year} {1983})}\BibitemShut {NoStop}%
\bibitem [{\citenamefont {Sikivie}(1983)}]{Sikivie:1983ip}%
  \BibitemOpen
  \bibfield  {author} {\bibinfo {author} {\bibfnamefont {P.}~\bibnamefont
  {Sikivie}},\ }\bibfield  {title} {\bibinfo {title} {{Experimental Tests of
  the Invisible Axion}},\ }\href {https://doi.org/10.1103/PhysRevLett.51.1415}
  {\bibfield  {journal} {\bibinfo  {journal} {Phys. Rev. Lett.}\ }\textbf
  {\bibinfo {volume} {51}},\ \bibinfo {pages} {1415} (\bibinfo {year}
  {1983})},\ \bibinfo {note} {[Erratum: Phys.Rev.Lett. 52, 695
  (1984)]}\BibitemShut {NoStop}%
\bibitem [{\citenamefont {Georgi}\ \emph {et~al.}(1974)\citenamefont {Georgi},
  \citenamefont {Quinn},\ and\ \citenamefont {Weinberg}}]{Georgi:1974yf}%
  \BibitemOpen
  \bibfield  {author} {\bibinfo {author} {\bibfnamefont {H.}~\bibnamefont
  {Georgi}}, \bibinfo {author} {\bibfnamefont {H.~R.}\ \bibnamefont {Quinn}},\
  and\ \bibinfo {author} {\bibfnamefont {S.}~\bibnamefont {Weinberg}},\
  }\bibfield  {title} {\bibinfo {title} {{Hierarchy of Interactions in Unified
  Gauge Theories}},\ }\href {https://doi.org/10.1103/PhysRevLett.33.451}
  {\bibfield  {journal} {\bibinfo  {journal} {Phys. Rev. Lett.}\ }\textbf
  {\bibinfo {volume} {33}},\ \bibinfo {pages} {451} (\bibinfo {year}
  {1974})}\BibitemShut {NoStop}%
\bibitem [{\citenamefont {Graham}\ \emph {et~al.}(2015)\citenamefont {Graham},
  \citenamefont {Kaplan},\ and\ \citenamefont {Rajendran}}]{Graham2015cka}%
  \BibitemOpen
  \bibfield  {author} {\bibinfo {author} {\bibfnamefont {P.~W.}\ \bibnamefont
  {Graham}}, \bibinfo {author} {\bibfnamefont {D.~E.}\ \bibnamefont {Kaplan}},\
  and\ \bibinfo {author} {\bibfnamefont {S.}~\bibnamefont {Rajendran}},\
  }\bibfield  {title} {\bibinfo {title} {{Cosmological Relaxation of the
  Electroweak Scale}},\ }\href {https://doi.org/10.1103/PhysRevLett.115.221801}
  {\bibfield  {journal} {\bibinfo  {journal} {Phys. Rev. Lett.}\ }\textbf
  {\bibinfo {volume} {115}},\ \bibinfo {pages} {221801} (\bibinfo {year}
  {2015})},\ \Eprint {https://arxiv.org/abs/1504.07551} {arXiv:1504.07551
  [hep-ph]} \BibitemShut {NoStop}%
\bibitem [{\citenamefont {Dine}\ and\ \citenamefont
  {Fischler}(1983)}]{dine1983not}%
  \BibitemOpen
  \bibfield  {author} {\bibinfo {author} {\bibfnamefont {M.}~\bibnamefont
  {Dine}}\ and\ \bibinfo {author} {\bibfnamefont {W.}~\bibnamefont
  {Fischler}},\ }\bibfield  {title} {\bibinfo {title} {The not-so-harmless
  axion},\ }\href@noop {} {\bibfield  {journal} {\bibinfo  {journal} {Physics
  Letters B}\ }\textbf {\bibinfo {volume} {120}},\ \bibinfo {pages} {137}
  (\bibinfo {year} {1983})}\BibitemShut {NoStop}%
\bibitem [{\citenamefont {Bertone}\ and\ \citenamefont
  {Hooper}(2018)}]{Bertone:2016nfn}%
  \BibitemOpen
  \bibfield  {author} {\bibinfo {author} {\bibfnamefont {G.}~\bibnamefont
  {Bertone}}\ and\ \bibinfo {author} {\bibfnamefont {D.}~\bibnamefont
  {Hooper}},\ }\bibfield  {title} {\bibinfo {title} {{History of dark
  matter}},\ }\href {https://doi.org/10.1103/RevModPhys.90.045002} {\bibfield
  {journal} {\bibinfo  {journal} {Rev. Mod. Phys.}\ }\textbf {\bibinfo {volume}
  {90}},\ \bibinfo {pages} {045002} (\bibinfo {year} {2018})},\ \Eprint
  {https://arxiv.org/abs/1605.04909} {arXiv:1605.04909 [astro-ph.CO]}
  \BibitemShut {NoStop}%
\bibitem [{\citenamefont {Freese}\ \emph {et~al.}(1990)\citenamefont {Freese},
  \citenamefont {Frieman},\ and\ \citenamefont {Olinto}}]{Freese1990rb}%
  \BibitemOpen
  \bibfield  {author} {\bibinfo {author} {\bibfnamefont {K.}~\bibnamefont
  {Freese}}, \bibinfo {author} {\bibfnamefont {J.~A.}\ \bibnamefont
  {Frieman}},\ and\ \bibinfo {author} {\bibfnamefont {A.~V.}\ \bibnamefont
  {Olinto}},\ }\bibfield  {title} {\bibinfo {title} {{Natural inflation with
  pseudo - Nambu-Goldstone bosons}},\ }\href
  {https://doi.org/10.1103/PhysRevLett.65.3233} {\bibfield  {journal} {\bibinfo
   {journal} {Phys. Rev. Lett.}\ }\textbf {\bibinfo {volume} {65}},\ \bibinfo
  {pages} {3233} (\bibinfo {year} {1990})}\BibitemShut {NoStop}%
\bibitem [{\citenamefont {Arvanitaki}\ \emph {et~al.}(2010)\citenamefont
  {Arvanitaki}, \citenamefont {Dimopoulos}, \citenamefont {Dubovsky},
  \citenamefont {Kaloper},\ and\ \citenamefont
  {March-Russell}}]{Arvanitaki:2009fg}%
  \BibitemOpen
  \bibfield  {author} {\bibinfo {author} {\bibfnamefont {A.}~\bibnamefont
  {Arvanitaki}}, \bibinfo {author} {\bibfnamefont {S.}~\bibnamefont
  {Dimopoulos}}, \bibinfo {author} {\bibfnamefont {S.}~\bibnamefont
  {Dubovsky}}, \bibinfo {author} {\bibfnamefont {N.}~\bibnamefont {Kaloper}},\
  and\ \bibinfo {author} {\bibfnamefont {J.}~\bibnamefont {March-Russell}},\
  }\bibfield  {title} {\bibinfo {title} {{String Axiverse}},\ }\href
  {https://doi.org/10.1103/PhysRevD.81.123530} {\bibfield  {journal} {\bibinfo
  {journal} {Phys. Rev. D}\ }\textbf {\bibinfo {volume} {81}},\ \bibinfo
  {pages} {123530} (\bibinfo {year} {2010})},\ \Eprint
  {https://arxiv.org/abs/0905.4720} {arXiv:0905.4720 [hep-th]} \BibitemShut
  {NoStop}%
\bibitem [{\citenamefont {Machado}\ \emph {et~al.}(2019)\citenamefont
  {Machado}, \citenamefont {Ratzinger}, \citenamefont {Schwaller},\ and\
  \citenamefont {Stefanek}}]{Machado:2018nqk}%
  \BibitemOpen
  \bibfield  {author} {\bibinfo {author} {\bibfnamefont {C.~S.}\ \bibnamefont
  {Machado}}, \bibinfo {author} {\bibfnamefont {W.}~\bibnamefont {Ratzinger}},
  \bibinfo {author} {\bibfnamefont {P.}~\bibnamefont {Schwaller}},\ and\
  \bibinfo {author} {\bibfnamefont {B.~A.}\ \bibnamefont {Stefanek}},\
  }\bibfield  {title} {\bibinfo {title} {{Audible Axions}},\ }\href
  {https://doi.org/10.1007/JHEP01(2019)053} {\bibfield  {journal} {\bibinfo
  {journal} {JHEP}\ }\textbf {\bibinfo {volume} {01}},\ \bibinfo {pages}
  {053}},\ \Eprint {https://arxiv.org/abs/1811.01950} {arXiv:1811.01950
  [hep-ph]} \BibitemShut {NoStop}%
\bibitem [{\citenamefont {Madge}\ \emph {et~al.}(2022)\citenamefont {Madge},
  \citenamefont {Ratzinger}, \citenamefont {Schmitt},\ and\ \citenamefont
  {Schwaller}}]{Madge:2021abk}%
  \BibitemOpen
  \bibfield  {author} {\bibinfo {author} {\bibfnamefont {E.}~\bibnamefont
  {Madge}}, \bibinfo {author} {\bibfnamefont {W.}~\bibnamefont {Ratzinger}},
  \bibinfo {author} {\bibfnamefont {D.}~\bibnamefont {Schmitt}},\ and\ \bibinfo
  {author} {\bibfnamefont {P.}~\bibnamefont {Schwaller}},\ }\bibfield  {title}
  {\bibinfo {title} {{Audible axions with a booster: Stochastic gravitational
  waves from rotating ALPs}},\ }\href
  {https://doi.org/10.21468/SciPostPhys.12.5.171} {\bibfield  {journal}
  {\bibinfo  {journal} {SciPost Phys.}\ }\textbf {\bibinfo {volume} {12}},\
  \bibinfo {pages} {171} (\bibinfo {year} {2022})},\ \Eprint
  {https://arxiv.org/abs/2111.12730} {arXiv:2111.12730 [hep-ph]} \BibitemShut
  {NoStop}%
\bibitem [{\citenamefont {Lue}\ \emph {et~al.}(1999)\citenamefont {Lue},
  \citenamefont {Wang},\ and\ \citenamefont {Kamionkowski}}]{Lue:1998mq}%
  \BibitemOpen
  \bibfield  {author} {\bibinfo {author} {\bibfnamefont {A.}~\bibnamefont
  {Lue}}, \bibinfo {author} {\bibfnamefont {L.-M.}\ \bibnamefont {Wang}},\ and\
  \bibinfo {author} {\bibfnamefont {M.}~\bibnamefont {Kamionkowski}},\
  }\bibfield  {title} {\bibinfo {title} {{Cosmological signature of new parity
  violating interactions}},\ }\href
  {https://doi.org/10.1103/PhysRevLett.83.1506} {\bibfield  {journal} {\bibinfo
   {journal} {Phys. Rev. Lett.}\ }\textbf {\bibinfo {volume} {83}},\ \bibinfo
  {pages} {1506} (\bibinfo {year} {1999})},\ \Eprint
  {https://arxiv.org/abs/astro-ph/9812088} {arXiv:astro-ph/9812088}
  \BibitemShut {NoStop}%
\bibitem [{\citenamefont {Nagano}\ \emph {et~al.}(2019)\citenamefont {Nagano},
  \citenamefont {Fujita}, \citenamefont {Michimura},\ and\ \citenamefont
  {Obata}}]{Nagano:2019rbw}%
  \BibitemOpen
  \bibfield  {author} {\bibinfo {author} {\bibfnamefont {K.}~\bibnamefont
  {Nagano}}, \bibinfo {author} {\bibfnamefont {T.}~\bibnamefont {Fujita}},
  \bibinfo {author} {\bibfnamefont {Y.}~\bibnamefont {Michimura}},\ and\
  \bibinfo {author} {\bibfnamefont {I.}~\bibnamefont {Obata}},\ }\bibfield
  {title} {\bibinfo {title} {{Axion Dark Matter Search with Interferometric
  Gravitational Wave Detectors}},\ }\href
  {https://doi.org/10.1103/PhysRevLett.123.111301} {\bibfield  {journal}
  {\bibinfo  {journal} {Phys. Rev. Lett.}\ }\textbf {\bibinfo {volume} {123}},\
  \bibinfo {pages} {111301} (\bibinfo {year} {2019})},\ \Eprint
  {https://arxiv.org/abs/1903.02017} {arXiv:1903.02017 [hep-ph]} \BibitemShut
  {NoStop}%
\bibitem [{\citenamefont {Heinze}\ \emph {et~al.}(2024)\citenamefont {Heinze},
  \citenamefont {Gill}, \citenamefont {Dmitriev}, \citenamefont {Smetana},
  \citenamefont {Yan}, \citenamefont {Boyer}, \citenamefont {Martynov},\ and\
  \citenamefont {Evans}}]{Heinze:2023nfb}%
  \BibitemOpen
  \bibfield  {author} {\bibinfo {author} {\bibfnamefont {J.}~\bibnamefont
  {Heinze}}, \bibinfo {author} {\bibfnamefont {A.}~\bibnamefont {Gill}},
  \bibinfo {author} {\bibfnamefont {A.}~\bibnamefont {Dmitriev}}, \bibinfo
  {author} {\bibfnamefont {J.}~\bibnamefont {Smetana}}, \bibinfo {author}
  {\bibfnamefont {T.}~\bibnamefont {Yan}}, \bibinfo {author} {\bibfnamefont
  {V.}~\bibnamefont {Boyer}}, \bibinfo {author} {\bibfnamefont
  {D.}~\bibnamefont {Martynov}},\ and\ \bibinfo {author} {\bibfnamefont
  {M.}~\bibnamefont {Evans}},\ }\bibfield  {title} {\bibinfo {title} {{First
  Results of the Laser-Interferometric Detector for Axions (LIDA)}},\ }\href
  {https://doi.org/10.1103/PhysRevLett.132.191002} {\bibfield  {journal}
  {\bibinfo  {journal} {Phys. Rev. Lett.}\ }\textbf {\bibinfo {volume} {132}},\
  \bibinfo {pages} {191002} (\bibinfo {year} {2024})},\ \Eprint
  {https://arxiv.org/abs/2307.01365} {arXiv:2307.01365 [astro-ph.CO]}
  \BibitemShut {NoStop}%
\bibitem [{\citenamefont {Yao}\ and\ \citenamefont {Tang}(2024)}]{Yao:2024fie}%
  \BibitemOpen
  \bibfield  {author} {\bibinfo {author} {\bibfnamefont {Y.-H.}\ \bibnamefont
  {Yao}}\ and\ \bibinfo {author} {\bibfnamefont {Y.}~\bibnamefont {Tang}},\
  }\bibfield  {title} {\bibinfo {title} {{Probing stochastic ultralight dark
  matter with space-based gravitational-wave interferometers}},\ }\href
  {https://doi.org/10.1103/PhysRevD.110.095015} {\bibfield  {journal} {\bibinfo
   {journal} {Phys. Rev. D}\ }\textbf {\bibinfo {volume} {110}},\ \bibinfo
  {pages} {095015} (\bibinfo {year} {2024})},\ \Eprint
  {https://arxiv.org/abs/2404.01494} {arXiv:2404.01494 [hep-ph]} \BibitemShut
  {NoStop}%
\bibitem [{\citenamefont {Gu\'e}\ \emph {et~al.}(2025)\citenamefont {Gu\'e},
  \citenamefont {Hees},\ and\ \citenamefont {Wolf}}]{Gue:2024txz}%
  \BibitemOpen
  \bibfield  {author} {\bibinfo {author} {\bibfnamefont {J.}~\bibnamefont
  {Gu\'e}}, \bibinfo {author} {\bibfnamefont {A.}~\bibnamefont {Hees}},\ and\
  \bibinfo {author} {\bibfnamefont {P.}~\bibnamefont {Wolf}},\ }\bibfield
  {title} {\bibinfo {title} {{Probing the axion\textendash{}photon coupling
  with space-based gravitational wave detectors}},\ }\href
  {https://doi.org/10.1088/1361-6382/adb23c} {\bibfield  {journal} {\bibinfo
  {journal} {Class. Quant. Grav.}\ }\textbf {\bibinfo {volume} {42}},\ \bibinfo
  {pages} {055015} (\bibinfo {year} {2025})},\ \Eprint
  {https://arxiv.org/abs/2410.17763} {arXiv:2410.17763 [hep-ph]} \BibitemShut
  {NoStop}%
\bibitem [{\citenamefont {Yao}\ \emph {et~al.}(2024)\citenamefont {Yao},
  \citenamefont {Jiang},\ and\ \citenamefont {Tang}}]{Yao:2024hap}%
  \BibitemOpen
  \bibfield  {author} {\bibinfo {author} {\bibfnamefont {Y.-H.}\ \bibnamefont
  {Yao}}, \bibinfo {author} {\bibfnamefont {T.}~\bibnamefont {Jiang}},\ and\
  \bibinfo {author} {\bibfnamefont {Y.}~\bibnamefont {Tang}},\ }\bibfield
  {title} {\bibinfo {title} {{Prospects for Axion Dark Matter Searches at
  LISA-like Interferometers}},\ }\href@noop {} {\bibfield  {journal} {\bibinfo
  {journal} {arXiv preprint}\ } (\bibinfo {year} {2024})},\ \Eprint
  {https://arxiv.org/abs/2410.22072} {arXiv:2410.22072 [hep-ph]} \BibitemShut
  {NoStop}%
\bibitem [{\citenamefont {Christensen}(2019)}]{Christensen:2018iqi}%
  \BibitemOpen
  \bibfield  {author} {\bibinfo {author} {\bibfnamefont {N.}~\bibnamefont
  {Christensen}},\ }\bibfield  {title} {\bibinfo {title} {{Stochastic
  Gravitational Wave Backgrounds}},\ }\href
  {https://doi.org/10.1088/1361-6633/aae6b5} {\bibfield  {journal} {\bibinfo
  {journal} {Rept. Prog. Phys.}\ }\textbf {\bibinfo {volume} {82}},\ \bibinfo
  {pages} {016903} (\bibinfo {year} {2019})},\ \Eprint
  {https://arxiv.org/abs/1811.08797} {arXiv:1811.08797 [gr-qc]} \BibitemShut
  {NoStop}%
\bibitem [{\citenamefont {Allen}\ and\ \citenamefont
  {Romano}(1999)}]{Allen:1997ad}%
  \BibitemOpen
  \bibfield  {author} {\bibinfo {author} {\bibfnamefont {B.}~\bibnamefont
  {Allen}}\ and\ \bibinfo {author} {\bibfnamefont {J.~D.}\ \bibnamefont
  {Romano}},\ }\bibfield  {title} {\bibinfo {title} {{Detecting a stochastic
  background of gravitational radiation: Signal processing strategies and
  sensitivities}},\ }\href {https://doi.org/10.1103/PhysRevD.59.102001}
  {\bibfield  {journal} {\bibinfo  {journal} {Phys. Rev. D}\ }\textbf {\bibinfo
  {volume} {59}},\ \bibinfo {pages} {102001} (\bibinfo {year} {1999})},\
  \Eprint {https://arxiv.org/abs/gr-qc/9710117} {arXiv:gr-qc/9710117}
  \BibitemShut {NoStop}%
\bibitem [{\citenamefont {Jackiw}\ and\ \citenamefont
  {Pi}(2003)}]{Jackiw:2003pm}%
  \BibitemOpen
  \bibfield  {author} {\bibinfo {author} {\bibfnamefont {R.}~\bibnamefont
  {Jackiw}}\ and\ \bibinfo {author} {\bibfnamefont {S.~Y.}\ \bibnamefont
  {Pi}},\ }\bibfield  {title} {\bibinfo {title} {{Chern-Simons modification of
  general relativity}},\ }\href {https://doi.org/10.1103/PhysRevD.68.104012}
  {\bibfield  {journal} {\bibinfo  {journal} {Phys. Rev. D}\ }\textbf {\bibinfo
  {volume} {68}},\ \bibinfo {pages} {104012} (\bibinfo {year} {2003})},\
  \Eprint {https://arxiv.org/abs/gr-qc/0308071} {arXiv:gr-qc/0308071}
  \BibitemShut {NoStop}%
\bibitem [{\citenamefont {Alexander}\ and\ \citenamefont
  {Yunes}(2009)}]{Alexander:2009tp}%
  \BibitemOpen
  \bibfield  {author} {\bibinfo {author} {\bibfnamefont {S.}~\bibnamefont
  {Alexander}}\ and\ \bibinfo {author} {\bibfnamefont {N.}~\bibnamefont
  {Yunes}},\ }\bibfield  {title} {\bibinfo {title} {{Chern-Simons Modified
  General Relativity}},\ }\href {https://doi.org/10.1016/j.physrep.2009.07.002}
  {\bibfield  {journal} {\bibinfo  {journal} {Phys. Rept.}\ }\textbf {\bibinfo
  {volume} {480}},\ \bibinfo {pages} {1} (\bibinfo {year} {2009})},\ \Eprint
  {https://arxiv.org/abs/0907.2562} {arXiv:0907.2562 [hep-th]} \BibitemShut
  {NoStop}%
\bibitem [{\citenamefont {Li}\ \emph {et~al.}(2020)\citenamefont {Li},
  \citenamefont {Rao},\ and\ \citenamefont {Zhao}}]{Li:2020xjt}%
  \BibitemOpen
  \bibfield  {author} {\bibinfo {author} {\bibfnamefont {M.}~\bibnamefont
  {Li}}, \bibinfo {author} {\bibfnamefont {H.}~\bibnamefont {Rao}},\ and\
  \bibinfo {author} {\bibfnamefont {D.}~\bibnamefont {Zhao}},\ }\bibfield
  {title} {\bibinfo {title} {{A simple parity violating gravity model without
  ghost instability}},\ }\href {https://doi.org/10.1088/1475-7516/2020/11/023}
  {\bibfield  {journal} {\bibinfo  {journal} {JCAP}\ }\textbf {\bibinfo
  {volume} {11}},\ \bibinfo {pages} {023}},\ \Eprint
  {https://arxiv.org/abs/2007.08038} {arXiv:2007.08038 [gr-qc]} \BibitemShut
  {NoStop}%
\bibitem [{\citenamefont {Li}\ \emph {et~al.}(2021)\citenamefont {Li},
  \citenamefont {Rao},\ and\ \citenamefont {Tong}}]{Li:2021wij}%
  \BibitemOpen
  \bibfield  {author} {\bibinfo {author} {\bibfnamefont {M.}~\bibnamefont
  {Li}}, \bibinfo {author} {\bibfnamefont {H.}~\bibnamefont {Rao}},\ and\
  \bibinfo {author} {\bibfnamefont {Y.}~\bibnamefont {Tong}},\ }\bibfield
  {title} {\bibinfo {title} {{Revisiting a parity violating gravity model
  without ghost instability: Local Lorentz covariance}},\ }\href
  {https://doi.org/10.1103/PhysRevD.104.084077} {\bibfield  {journal} {\bibinfo
   {journal} {Phys. Rev. D}\ }\textbf {\bibinfo {volume} {104}},\ \bibinfo
  {pages} {084077} (\bibinfo {year} {2021})},\ \Eprint
  {https://arxiv.org/abs/2104.05917} {arXiv:2104.05917 [gr-qc]} \BibitemShut
  {NoStop}%
\bibitem [{\citenamefont {Cai}\ \emph {et~al.}(2022)\citenamefont {Cai},
  \citenamefont {Fu},\ and\ \citenamefont {Yu}}]{Cai:2021uup}%
  \BibitemOpen
  \bibfield  {author} {\bibinfo {author} {\bibfnamefont {R.-G.}\ \bibnamefont
  {Cai}}, \bibinfo {author} {\bibfnamefont {C.}~\bibnamefont {Fu}},\ and\
  \bibinfo {author} {\bibfnamefont {W.-W.}\ \bibnamefont {Yu}},\ }\bibfield
  {title} {\bibinfo {title} {{Parity violation in stochastic gravitational wave
  background from inflation in Nieh-Yan modified teleparallel gravity}},\
  }\href {https://doi.org/10.1103/PhysRevD.105.103520} {\bibfield  {journal}
  {\bibinfo  {journal} {Phys. Rev. D}\ }\textbf {\bibinfo {volume} {105}},\
  \bibinfo {pages} {103520} (\bibinfo {year} {2022})},\ \Eprint
  {https://arxiv.org/abs/2112.04794} {arXiv:2112.04794 [astro-ph.CO]}
  \BibitemShut {NoStop}%
\bibitem [{\citenamefont {Wu}\ \emph {et~al.}(2022)\citenamefont {Wu},
  \citenamefont {Zhu}, \citenamefont {Niu}, \citenamefont {Zhao},\ and\
  \citenamefont {Wang}}]{Wu:2021ndf}%
  \BibitemOpen
  \bibfield  {author} {\bibinfo {author} {\bibfnamefont {Q.}~\bibnamefont
  {Wu}}, \bibinfo {author} {\bibfnamefont {T.}~\bibnamefont {Zhu}}, \bibinfo
  {author} {\bibfnamefont {R.}~\bibnamefont {Niu}}, \bibinfo {author}
  {\bibfnamefont {W.}~\bibnamefont {Zhao}},\ and\ \bibinfo {author}
  {\bibfnamefont {A.}~\bibnamefont {Wang}},\ }\bibfield  {title} {\bibinfo
  {title} {{Constraints on the Nieh-Yan modified teleparallel gravity with
  gravitational waves}},\ }\href {https://doi.org/10.1103/PhysRevD.105.024035}
  {\bibfield  {journal} {\bibinfo  {journal} {Phys. Rev. D}\ }\textbf {\bibinfo
  {volume} {105}},\ \bibinfo {pages} {024035} (\bibinfo {year} {2022})},\
  \Eprint {https://arxiv.org/abs/2110.13870} {arXiv:2110.13870 [gr-qc]}
  \BibitemShut {NoStop}%
\bibitem [{\citenamefont {Li}\ and\ \citenamefont {Rao}(2023)}]{Li:2023fto}%
  \BibitemOpen
  \bibfield  {author} {\bibinfo {author} {\bibfnamefont {M.}~\bibnamefont
  {Li}}\ and\ \bibinfo {author} {\bibfnamefont {H.}~\bibnamefont {Rao}},\
  }\bibfield  {title} {\bibinfo {title} {{Irregular universe in the Nieh-Yan
  modified teleparallel gravity}},\ }\href
  {https://doi.org/10.1016/j.physletb.2023.137929} {\bibfield  {journal}
  {\bibinfo  {journal} {Phys. Lett. B}\ }\textbf {\bibinfo {volume} {841}},\
  \bibinfo {pages} {137929} (\bibinfo {year} {2023})},\ \Eprint
  {https://arxiv.org/abs/2301.02847} {arXiv:2301.02847 [gr-qc]} \BibitemShut
  {NoStop}%
\bibitem [{\citenamefont {Li}\ \emph {et~al.}(2022)\citenamefont {Li},
  \citenamefont {Tong},\ and\ \citenamefont {Zhao}}]{Li:2022vtn}%
  \BibitemOpen
  \bibfield  {author} {\bibinfo {author} {\bibfnamefont {M.}~\bibnamefont
  {Li}}, \bibinfo {author} {\bibfnamefont {Y.}~\bibnamefont {Tong}},\ and\
  \bibinfo {author} {\bibfnamefont {D.}~\bibnamefont {Zhao}},\ }\bibfield
  {title} {\bibinfo {title} {{Possible consistent model of parity violations in
  the symmetric teleparallel gravity}},\ }\href
  {https://doi.org/10.1103/PhysRevD.105.104002} {\bibfield  {journal} {\bibinfo
   {journal} {Phys. Rev. D}\ }\textbf {\bibinfo {volume} {105}},\ \bibinfo
  {pages} {104002} (\bibinfo {year} {2022})},\ \Eprint
  {https://arxiv.org/abs/2203.06912} {arXiv:2203.06912 [gr-qc]} \BibitemShut
  {NoStop}%
\bibitem [{\citenamefont {Rao}\ and\ \citenamefont {Zhao}(2023)}]{Rao:2023doc}%
  \BibitemOpen
  \bibfield  {author} {\bibinfo {author} {\bibfnamefont {H.}~\bibnamefont
  {Rao}}\ and\ \bibinfo {author} {\bibfnamefont {D.}~\bibnamefont {Zhao}},\
  }\bibfield  {title} {\bibinfo {title} {{Parity violating scalar-tensor model
  in teleparallel gravity and its cosmological application}},\ }\href
  {https://doi.org/10.1007/JHEP08(2023)070} {\bibfield  {journal} {\bibinfo
  {journal} {JHEP}\ }\textbf {\bibinfo {volume} {08}},\ \bibinfo {pages}
  {070}},\ \Eprint {https://arxiv.org/abs/2304.07138} {arXiv:2304.07138
  [gr-qc]} \BibitemShut {NoStop}%
\bibitem [{\citenamefont {Zhang}\ \emph {et~al.}(2024)\citenamefont {Zhang},
  \citenamefont {Feng},\ and\ \citenamefont {Gao}}]{Zhang:2024vfw}%
  \BibitemOpen
  \bibfield  {author} {\bibinfo {author} {\bibfnamefont {F.}~\bibnamefont
  {Zhang}}, \bibinfo {author} {\bibfnamefont {J.-X.}\ \bibnamefont {Feng}},\
  and\ \bibinfo {author} {\bibfnamefont {X.}~\bibnamefont {Gao}},\ }\bibfield
  {title} {\bibinfo {title} {{Scalar induced gravitational waves in metric
  teleparallel gravity with the Nieh-Yan term}},\ }\href
  {https://doi.org/10.1103/PhysRevD.110.023537} {\bibfield  {journal} {\bibinfo
   {journal} {Phys. Rev. D}\ }\textbf {\bibinfo {volume} {110}},\ \bibinfo
  {pages} {023537} (\bibinfo {year} {2024})},\ \Eprint
  {https://arxiv.org/abs/2404.02922} {arXiv:2404.02922 [gr-qc]} \BibitemShut
  {NoStop}%
\bibitem [{\citenamefont {Xu}\ \emph {et~al.}(2024)\citenamefont {Xu},
  \citenamefont {Ding}, \citenamefont {Su}, \citenamefont {Chen},\ and\
  \citenamefont {Zhang}}]{Xu:2024kwy}%
  \BibitemOpen
  \bibfield  {author} {\bibinfo {author} {\bibfnamefont {B.}~\bibnamefont
  {Xu}}, \bibinfo {author} {\bibfnamefont {K.}~\bibnamefont {Ding}}, \bibinfo
  {author} {\bibfnamefont {H.}~\bibnamefont {Su}}, \bibinfo {author}
  {\bibfnamefont {J.}~\bibnamefont {Chen}},\ and\ \bibinfo {author}
  {\bibfnamefont {Y.-L.}\ \bibnamefont {Zhang}},\ }\bibfield  {title} {\bibinfo
  {title} {{Chiral Gravitational Wave Background from Audible Axion via
  Nieh-Yan Term}},\ }\href@noop {} {\bibfield  {journal} {\bibinfo  {journal}
  {arXiv preprint}\ } (\bibinfo {year} {2024})},\ \Eprint
  {https://arxiv.org/abs/2411.08691} {arXiv:2411.08691 [hep-ph]} \BibitemShut
  {NoStop}%
\bibitem [{\citenamefont {Seto}\ and\ \citenamefont
  {Taruya}(2007)}]{Seto:2007tn}%
  \BibitemOpen
  \bibfield  {author} {\bibinfo {author} {\bibfnamefont {N.}~\bibnamefont
  {Seto}}\ and\ \bibinfo {author} {\bibfnamefont {A.}~\bibnamefont {Taruya}},\
  }\bibfield  {title} {\bibinfo {title} {{Measuring a Parity Violation
  Signature in the Early Universe via Ground-based Laser Interferometers}},\
  }\href {https://doi.org/10.1103/PhysRevLett.99.121101} {\bibfield  {journal}
  {\bibinfo  {journal} {Phys. Rev. Lett.}\ }\textbf {\bibinfo {volume} {99}},\
  \bibinfo {pages} {121101} (\bibinfo {year} {2007})},\ \Eprint
  {https://arxiv.org/abs/0707.0535} {arXiv:0707.0535 [astro-ph]} \BibitemShut
  {NoStop}%
\bibitem [{\citenamefont {Seto}\ and\ \citenamefont
  {Taruya}(2008)}]{Seto:2008sr}%
  \BibitemOpen
  \bibfield  {author} {\bibinfo {author} {\bibfnamefont {N.}~\bibnamefont
  {Seto}}\ and\ \bibinfo {author} {\bibfnamefont {A.}~\bibnamefont {Taruya}},\
  }\bibfield  {title} {\bibinfo {title} {{Polarization analysis of
  gravitational-wave backgrounds from the correlation signals of ground-based
  interferometers: Measuring a circular-polarization mode}},\ }\href
  {https://doi.org/10.1103/PhysRevD.77.103001} {\bibfield  {journal} {\bibinfo
  {journal} {Phys. Rev. D}\ }\textbf {\bibinfo {volume} {77}},\ \bibinfo
  {pages} {103001} (\bibinfo {year} {2008})},\ \Eprint
  {https://arxiv.org/abs/0801.4185} {arXiv:0801.4185 [astro-ph]} \BibitemShut
  {NoStop}%
\bibitem [{\citenamefont {Smith}\ and\ \citenamefont
  {Caldwell}(2017)}]{Smith:2016jqs}%
  \BibitemOpen
  \bibfield  {author} {\bibinfo {author} {\bibfnamefont {T.~L.}\ \bibnamefont
  {Smith}}\ and\ \bibinfo {author} {\bibfnamefont {R.}~\bibnamefont
  {Caldwell}},\ }\bibfield  {title} {\bibinfo {title} {{Sensitivity to a
  Frequency-Dependent Circular Polarization in an Isotropic Stochastic
  Gravitational Wave Background}},\ }\href
  {https://doi.org/10.1103/PhysRevD.95.044036} {\bibfield  {journal} {\bibinfo
  {journal} {Phys. Rev. D}\ }\textbf {\bibinfo {volume} {95}},\ \bibinfo
  {pages} {044036} (\bibinfo {year} {2017})},\ \Eprint
  {https://arxiv.org/abs/1609.05901} {arXiv:1609.05901 [gr-qc]} \BibitemShut
  {NoStop}%
\bibitem [{\citenamefont {Seto}(2006{\natexlab{a}})}]{Seto:2006hf}%
  \BibitemOpen
  \bibfield  {author} {\bibinfo {author} {\bibfnamefont {N.}~\bibnamefont
  {Seto}},\ }\bibfield  {title} {\bibinfo {title} {{Prospects for direct
  detection of circular polarization of gravitational-wave background}},\
  }\href {https://doi.org/10.1103/PhysRevLett.97.151101} {\bibfield  {journal}
  {\bibinfo  {journal} {Phys. Rev. Lett.}\ }\textbf {\bibinfo {volume} {97}},\
  \bibinfo {pages} {151101} (\bibinfo {year} {2006}{\natexlab{a}})},\ \Eprint
  {https://arxiv.org/abs/astro-ph/0609504} {arXiv:astro-ph/0609504}
  \BibitemShut {NoStop}%
\bibitem [{\citenamefont {Orlando}\ \emph {et~al.}(2021)\citenamefont
  {Orlando}, \citenamefont {Pieroni},\ and\ \citenamefont
  {Ricciardone}}]{Orlando:2020oko}%
  \BibitemOpen
  \bibfield  {author} {\bibinfo {author} {\bibfnamefont {G.}~\bibnamefont
  {Orlando}}, \bibinfo {author} {\bibfnamefont {M.}~\bibnamefont {Pieroni}},\
  and\ \bibinfo {author} {\bibfnamefont {A.}~\bibnamefont {Ricciardone}},\
  }\bibfield  {title} {\bibinfo {title} {{Measuring Parity Violation in the
  Stochastic Gravitational Wave Background with the LISA-Taiji network}},\
  }\href {https://doi.org/10.1088/1475-7516/2021/03/069} {\bibfield  {journal}
  {\bibinfo  {journal} {JCAP}\ }\textbf {\bibinfo {volume} {03}},\ \bibinfo
  {pages} {069}},\ \Eprint {https://arxiv.org/abs/2011.07059} {arXiv:2011.07059
  [astro-ph.CO]} \BibitemShut {NoStop}%
\bibitem [{\citenamefont {Saito}\ \emph {et~al.}(2007)\citenamefont {Saito},
  \citenamefont {Ichiki},\ and\ \citenamefont {Taruya}}]{Saito:2007kt}%
  \BibitemOpen
  \bibfield  {author} {\bibinfo {author} {\bibfnamefont {S.}~\bibnamefont
  {Saito}}, \bibinfo {author} {\bibfnamefont {K.}~\bibnamefont {Ichiki}},\ and\
  \bibinfo {author} {\bibfnamefont {A.}~\bibnamefont {Taruya}},\ }\bibfield
  {title} {\bibinfo {title} {{Probing polarization states of primordial
  gravitational waves with CMB anisotropies}},\ }\href
  {https://doi.org/10.1088/1475-7516/2007/09/002} {\bibfield  {journal}
  {\bibinfo  {journal} {JCAP}\ }\textbf {\bibinfo {volume} {09}},\ \bibinfo
  {pages} {002}},\ \Eprint {https://arxiv.org/abs/0705.3701} {arXiv:0705.3701
  [astro-ph]} \BibitemShut {NoStop}%
\bibitem [{\citenamefont {Sorbo}(2011)}]{Sorbo:2011rz}%
  \BibitemOpen
  \bibfield  {author} {\bibinfo {author} {\bibfnamefont {L.}~\bibnamefont
  {Sorbo}},\ }\bibfield  {title} {\bibinfo {title} {{Parity violation in the
  Cosmic Microwave Background from a pseudoscalar inflaton}},\ }\href
  {https://doi.org/10.1088/1475-7516/2011/06/003} {\bibfield  {journal}
  {\bibinfo  {journal} {JCAP}\ }\textbf {\bibinfo {volume} {06}},\ \bibinfo
  {pages} {003}},\ \Eprint {https://arxiv.org/abs/1101.1525} {arXiv:1101.1525
  [astro-ph.CO]} \BibitemShut {NoStop}%
\bibitem [{\citenamefont {Amaro-Seoane}\ \emph {et~al.}(2017)\citenamefont
  {Amaro-Seoane} \emph {et~al.}}]{LISA:2017pwj}%
  \BibitemOpen
  \bibfield  {author} {\bibinfo {author} {\bibfnamefont {P.}~\bibnamefont
  {Amaro-Seoane}} \emph {et~al.} (\bibinfo {collaboration} {LISA}),\ }\bibfield
   {title} {\bibinfo {title} {{Laser Interferometer Space Antenna}},\
  }\href@noop {} {\bibfield  {journal} {\bibinfo  {journal} {arXiv preprint}\ }
  (\bibinfo {year} {2017})},\ \Eprint {https://arxiv.org/abs/1702.00786}
  {arXiv:1702.00786 [astro-ph.IM]} \BibitemShut {NoStop}%
\bibitem [{\citenamefont {Ruan}\ \emph
  {et~al.}(2020{\natexlab{a}})\citenamefont {Ruan}, \citenamefont {Guo},
  \citenamefont {Cai},\ and\ \citenamefont {Zhang}}]{Ruan:2018tsw}%
  \BibitemOpen
  \bibfield  {author} {\bibinfo {author} {\bibfnamefont {W.-H.}\ \bibnamefont
  {Ruan}}, \bibinfo {author} {\bibfnamefont {Z.-K.}\ \bibnamefont {Guo}},
  \bibinfo {author} {\bibfnamefont {R.-G.}\ \bibnamefont {Cai}},\ and\ \bibinfo
  {author} {\bibfnamefont {Y.-Z.}\ \bibnamefont {Zhang}},\ }\bibfield  {title}
  {\bibinfo {title} {{Taiji program: Gravitational-wave sources}},\ }\href
  {https://doi.org/10.1142/S0217751X2050075X} {\bibfield  {journal} {\bibinfo
  {journal} {Int. J. Mod. Phys. A}\ }\textbf {\bibinfo {volume} {35}},\
  \bibinfo {pages} {2050075} (\bibinfo {year} {2020}{\natexlab{a}})},\ \Eprint
  {https://arxiv.org/abs/1807.09495} {arXiv:1807.09495 [gr-qc]} \BibitemShut
  {NoStop}%
\bibitem [{\citenamefont {Seto}(2006{\natexlab{b}})}]{Seto:2005qy}%
  \BibitemOpen
  \bibfield  {author} {\bibinfo {author} {\bibfnamefont {N.}~\bibnamefont
  {Seto}},\ }\bibfield  {title} {\bibinfo {title} {{Correlation analysis of
  stochastic gravitational wave background around 0.1-1 Hz}},\ }\href
  {https://doi.org/10.1103/PhysRevD.73.063001} {\bibfield  {journal} {\bibinfo
  {journal} {Phys. Rev. D}\ }\textbf {\bibinfo {volume} {73}},\ \bibinfo
  {pages} {063001} (\bibinfo {year} {2006}{\natexlab{b}})},\ \Eprint
  {https://arxiv.org/abs/gr-qc/0510067} {arXiv:gr-qc/0510067} \BibitemShut
  {NoStop}%
\bibitem [{\citenamefont {Abbott}\ \emph {et~al.}(2007)\citenamefont {Abbott}
  \emph {et~al.}}]{LIGOScientific:2006zmq}%
  \BibitemOpen
  \bibfield  {author} {\bibinfo {author} {\bibfnamefont {B.}~\bibnamefont
  {Abbott}} \emph {et~al.} (\bibinfo {collaboration} {LIGO Scientific}),\
  }\bibfield  {title} {\bibinfo {title} {{Searching for a Stochastic Background
  of Gravitational Waves with LIGO}},\ }\href {https://doi.org/10.1086/511329}
  {\bibfield  {journal} {\bibinfo  {journal} {Astrophys. J.}\ }\textbf
  {\bibinfo {volume} {659}},\ \bibinfo {pages} {918} (\bibinfo {year}
  {2007})},\ \Eprint {https://arxiv.org/abs/astro-ph/0608606}
  {arXiv:astro-ph/0608606} \BibitemShut {NoStop}%
\bibitem [{\citenamefont {Schutz}(2011)}]{Schutz:2011tw}%
  \BibitemOpen
  \bibfield  {author} {\bibinfo {author} {\bibfnamefont {B.~F.}\ \bibnamefont
  {Schutz}},\ }\bibfield  {title} {\bibinfo {title} {{Networks of gravitational
  wave detectors and three figures of merit}},\ }\href
  {https://doi.org/10.1088/0264-9381/28/12/125023} {\bibfield  {journal}
  {\bibinfo  {journal} {Class. Quant. Grav.}\ }\textbf {\bibinfo {volume}
  {28}},\ \bibinfo {pages} {125023} (\bibinfo {year} {2011})},\ \Eprint
  {https://arxiv.org/abs/1102.5421} {arXiv:1102.5421 [astro-ph.IM]}
  \BibitemShut {NoStop}%
\bibitem [{\citenamefont {Seto}(2020)}]{Seto:2020zxw}%
  \BibitemOpen
  \bibfield  {author} {\bibinfo {author} {\bibfnamefont {N.}~\bibnamefont
  {Seto}},\ }\bibfield  {title} {\bibinfo {title} {{Measuring Parity Asymmetry
  of Gravitational Wave Backgrounds with a Heliocentric Detector Network in the
  mHz Band}},\ }\href {https://doi.org/10.1103/PhysRevLett.125.251101}
  {\bibfield  {journal} {\bibinfo  {journal} {Phys. Rev. Lett.}\ }\textbf
  {\bibinfo {volume} {125}},\ \bibinfo {pages} {251101} (\bibinfo {year}
  {2020})},\ \Eprint {https://arxiv.org/abs/2009.02928} {arXiv:2009.02928
  [gr-qc]} \BibitemShut {NoStop}%
\bibitem [{\citenamefont {Ruan}\ \emph
  {et~al.}(2020{\natexlab{b}})\citenamefont {Ruan}, \citenamefont {Liu},
  \citenamefont {Guo}, \citenamefont {Wu},\ and\ \citenamefont
  {Cai}}]{Ruan:2020smc}%
  \BibitemOpen
  \bibfield  {author} {\bibinfo {author} {\bibfnamefont {W.-H.}\ \bibnamefont
  {Ruan}}, \bibinfo {author} {\bibfnamefont {C.}~\bibnamefont {Liu}}, \bibinfo
  {author} {\bibfnamefont {Z.-K.}\ \bibnamefont {Guo}}, \bibinfo {author}
  {\bibfnamefont {Y.-L.}\ \bibnamefont {Wu}},\ and\ \bibinfo {author}
  {\bibfnamefont {R.-G.}\ \bibnamefont {Cai}},\ }\bibfield  {title} {\bibinfo
  {title} {{The LISA-Taiji network}},\ }\href
  {https://doi.org/10.1038/s41550-019-1008-4} {\bibfield  {journal} {\bibinfo
  {journal} {Nature Astron.}\ }\textbf {\bibinfo {volume} {4}},\ \bibinfo
  {pages} {108} (\bibinfo {year} {2020}{\natexlab{b}})},\ \Eprint
  {https://arxiv.org/abs/2002.03603} {arXiv:2002.03603 [gr-qc]} \BibitemShut
  {NoStop}%
\bibitem [{\citenamefont {Wang}\ \emph {et~al.}(2021)\citenamefont {Wang},
  \citenamefont {Ni}, \citenamefont {Han}, \citenamefont {Xu},\ and\
  \citenamefont {Luo}}]{Wang:2021uih}%
  \BibitemOpen
  \bibfield  {author} {\bibinfo {author} {\bibfnamefont {G.}~\bibnamefont
  {Wang}}, \bibinfo {author} {\bibfnamefont {W.-T.}\ \bibnamefont {Ni}},
  \bibinfo {author} {\bibfnamefont {W.-B.}\ \bibnamefont {Han}}, \bibinfo
  {author} {\bibfnamefont {P.}~\bibnamefont {Xu}},\ and\ \bibinfo {author}
  {\bibfnamefont {Z.}~\bibnamefont {Luo}},\ }\bibfield  {title} {\bibinfo
  {title} {{Alternative LISA-TAIJI networks}},\ }\href
  {https://doi.org/10.1103/PhysRevD.104.024012} {\bibfield  {journal} {\bibinfo
   {journal} {Phys. Rev. D}\ }\textbf {\bibinfo {volume} {104}},\ \bibinfo
  {pages} {024012} (\bibinfo {year} {2021})},\ \Eprint
  {https://arxiv.org/abs/2105.00746} {arXiv:2105.00746 [gr-qc]} \BibitemShut
  {NoStop}%
\bibitem [{\citenamefont {Wang}\ and\ \citenamefont
  {Han}(2021{\natexlab{a}})}]{Wang:2021njt}%
  \BibitemOpen
  \bibfield  {author} {\bibinfo {author} {\bibfnamefont {G.}~\bibnamefont
  {Wang}}\ and\ \bibinfo {author} {\bibfnamefont {W.-B.}\ \bibnamefont {Han}},\
  }\bibfield  {title} {\bibinfo {title} {{Alternative LISA-TAIJI networks:
  Detectability of the isotropic stochastic gravitational wave background}},\
  }\href {https://doi.org/10.1103/PhysRevD.104.104015} {\bibfield  {journal}
  {\bibinfo  {journal} {Phys. Rev. D}\ }\textbf {\bibinfo {volume} {104}},\
  \bibinfo {pages} {104015} (\bibinfo {year} {2021}{\natexlab{a}})},\ \Eprint
  {https://arxiv.org/abs/2108.11151} {arXiv:2108.11151 [gr-qc]} \BibitemShut
  {NoStop}%
\bibitem [{\citenamefont {Cai}\ \emph {et~al.}(2024)\citenamefont {Cai},
  \citenamefont {Guo}, \citenamefont {Hu}, \citenamefont {Liu}, \citenamefont
  {Lu}, \citenamefont {Ni}, \citenamefont {Ruan}, \citenamefont {Seto},
  \citenamefont {Wang},\ and\ \citenamefont {Wu}}]{Cai:2023ywp}%
  \BibitemOpen
  \bibfield  {author} {\bibinfo {author} {\bibfnamefont {R.-G.}\ \bibnamefont
  {Cai}}, \bibinfo {author} {\bibfnamefont {Z.-K.}\ \bibnamefont {Guo}},
  \bibinfo {author} {\bibfnamefont {B.}~\bibnamefont {Hu}}, \bibinfo {author}
  {\bibfnamefont {C.}~\bibnamefont {Liu}}, \bibinfo {author} {\bibfnamefont
  {Y.}~\bibnamefont {Lu}}, \bibinfo {author} {\bibfnamefont {W.-T.}\
  \bibnamefont {Ni}}, \bibinfo {author} {\bibfnamefont {W.-H.}\ \bibnamefont
  {Ruan}}, \bibinfo {author} {\bibfnamefont {N.}~\bibnamefont {Seto}}, \bibinfo
  {author} {\bibfnamefont {G.}~\bibnamefont {Wang}},\ and\ \bibinfo {author}
  {\bibfnamefont {Y.-L.}\ \bibnamefont {Wu}},\ }\bibfield  {title} {\bibinfo
  {title} {{On networks of space-based gravitational-wave detectors}},\ }\href
  {https://doi.org/10.1016/j.fmre.2023.10.007} {\bibfield  {journal} {\bibinfo
  {journal} {Fund. Res.}\ }\textbf {\bibinfo {volume} {4}},\ \bibinfo {pages}
  {1072} (\bibinfo {year} {2024})},\ \Eprint {https://arxiv.org/abs/2305.04551}
  {arXiv:2305.04551 [gr-qc]} \BibitemShut {NoStop}%
\bibitem [{\citenamefont {Zhao}\ and\ \citenamefont
  {Wang}(2024)}]{Zhao:2024yau}%
  \BibitemOpen
  \bibfield  {author} {\bibinfo {author} {\bibfnamefont {Z.-C.}\ \bibnamefont
  {Zhao}}\ and\ \bibinfo {author} {\bibfnamefont {S.}~\bibnamefont {Wang}},\
  }\bibfield  {title} {\bibinfo {title} {{Measuring the anisotropies in
  astrophysical and cosmological gravitational-wave backgrounds with Taiji and
  LISA networks}},\ }\href {https://doi.org/10.1007/s11433-024-2498-0}
  {\bibfield  {journal} {\bibinfo  {journal} {Sci. China Phys. Mech. Astron.}\
  }\textbf {\bibinfo {volume} {67}},\ \bibinfo {pages} {120411} (\bibinfo
  {year} {2024})},\ \Eprint {https://arxiv.org/abs/2407.09380}
  {arXiv:2407.09380 [gr-qc]} \BibitemShut {NoStop}%
\bibitem [{\citenamefont {Chen}\ \emph
  {et~al.}(2024{\natexlab{a}})\citenamefont {Chen}, \citenamefont {Liu},\ and\
  \citenamefont {Zhang}}]{Chen:2024ikn}%
  \BibitemOpen
  \bibfield  {author} {\bibinfo {author} {\bibfnamefont {J.}~\bibnamefont
  {Chen}}, \bibinfo {author} {\bibfnamefont {C.}~\bibnamefont {Liu}},\ and\
  \bibinfo {author} {\bibfnamefont {Y.-L.}\ \bibnamefont {Zhang}},\ }\bibfield
  {title} {\bibinfo {title} {{Parity-violating Gravitational Wave Background
  Search with a Network of Space-borne Triangular Detectors}},\ }\href@noop {}
  {\bibfield  {journal} {\bibinfo  {journal} {arXiv}\ } (\bibinfo {year}
  {2024}{\natexlab{a}})},\ \Eprint {https://arxiv.org/abs/2410.18916}
  {arXiv:2410.18916 [gr-qc]} \BibitemShut {NoStop}%
\bibitem [{\citenamefont {Wang}\ and\ \citenamefont
  {Han}(2021{\natexlab{b}})}]{Wang:2021mou}%
  \BibitemOpen
  \bibfield  {author} {\bibinfo {author} {\bibfnamefont {G.}~\bibnamefont
  {Wang}}\ and\ \bibinfo {author} {\bibfnamefont {W.-B.}\ \bibnamefont {Han}},\
  }\bibfield  {title} {\bibinfo {title} {{Observing gravitational wave
  polarizations with the LISA-TAIJI network}},\ }\href
  {https://doi.org/10.1103/PhysRevD.103.064021} {\bibfield  {journal} {\bibinfo
   {journal} {Phys. Rev. D}\ }\textbf {\bibinfo {volume} {103}},\ \bibinfo
  {pages} {064021} (\bibinfo {year} {2021}{\natexlab{b}})},\ \Eprint
  {https://arxiv.org/abs/2101.01991} {arXiv:2101.01991 [gr-qc]} \BibitemShut
  {NoStop}%
\bibitem [{\citenamefont {Zhang}\ \emph
  {et~al.}(2022{\natexlab{a}})\citenamefont {Zhang}, \citenamefont {Zhao},
  \citenamefont {Mohanty},\ and\ \citenamefont {Liu}}]{Zhang:2022wcp}%
  \BibitemOpen
  \bibfield  {author} {\bibinfo {author} {\bibfnamefont {X.-H.}\ \bibnamefont
  {Zhang}}, \bibinfo {author} {\bibfnamefont {S.-D.}\ \bibnamefont {Zhao}},
  \bibinfo {author} {\bibfnamefont {S.~D.}\ \bibnamefont {Mohanty}},\ and\
  \bibinfo {author} {\bibfnamefont {Y.-X.}\ \bibnamefont {Liu}},\ }\bibfield
  {title} {\bibinfo {title} {{Resolving Galactic binaries using a network of
  space-borne gravitational wave detectors}},\ }\href
  {https://doi.org/10.1103/PhysRevD.106.102004} {\bibfield  {journal} {\bibinfo
   {journal} {Phys. Rev. D}\ }\textbf {\bibinfo {volume} {106}},\ \bibinfo
  {pages} {102004} (\bibinfo {year} {2022}{\natexlab{a}})},\ \Eprint
  {https://arxiv.org/abs/2206.12083} {arXiv:2206.12083 [gr-qc]} \BibitemShut
  {NoStop}%
\bibitem [{\citenamefont {Zhang}\ \emph
  {et~al.}(2022{\natexlab{b}})\citenamefont {Zhang}, \citenamefont {Gong},\
  and\ \citenamefont {Zhang}}]{Zhang:2021wwd}%
  \BibitemOpen
  \bibfield  {author} {\bibinfo {author} {\bibfnamefont {C.}~\bibnamefont
  {Zhang}}, \bibinfo {author} {\bibfnamefont {Y.}~\bibnamefont {Gong}},\ and\
  \bibinfo {author} {\bibfnamefont {C.}~\bibnamefont {Zhang}},\ }\bibfield
  {title} {\bibinfo {title} {{Source localizations with the network of
  space-based gravitational wave detectors}},\ }\href
  {https://doi.org/10.1103/PhysRevD.106.024004} {\bibfield  {journal} {\bibinfo
   {journal} {Phys. Rev. D}\ }\textbf {\bibinfo {volume} {106}},\ \bibinfo
  {pages} {024004} (\bibinfo {year} {2022}{\natexlab{b}})},\ \Eprint
  {https://arxiv.org/abs/2112.02299} {arXiv:2112.02299 [gr-qc]} \BibitemShut
  {NoStop}%
\bibitem [{\citenamefont {Shuman}\ and\ \citenamefont
  {Cornish}(2022)}]{Shuman:2021ruh}%
  \BibitemOpen
  \bibfield  {author} {\bibinfo {author} {\bibfnamefont {K.~J.}\ \bibnamefont
  {Shuman}}\ and\ \bibinfo {author} {\bibfnamefont {N.~J.}\ \bibnamefont
  {Cornish}},\ }\bibfield  {title} {\bibinfo {title} {{Massive black hole
  binaries and where to find them with dual detector networks}},\ }\href
  {https://doi.org/10.1103/PhysRevD.105.064055} {\bibfield  {journal} {\bibinfo
   {journal} {Phys. Rev. D}\ }\textbf {\bibinfo {volume} {105}},\ \bibinfo
  {pages} {064055} (\bibinfo {year} {2022})},\ \Eprint
  {https://arxiv.org/abs/2105.02943} {arXiv:2105.02943 [gr-qc]} \BibitemShut
  {NoStop}%
\bibitem [{\citenamefont {Ruan}\ \emph {et~al.}(2021)\citenamefont {Ruan},
  \citenamefont {Liu}, \citenamefont {Guo}, \citenamefont {Wu},\ and\
  \citenamefont {Cai}}]{Ruan:2019tje}%
  \BibitemOpen
  \bibfield  {author} {\bibinfo {author} {\bibfnamefont {W.-H.}\ \bibnamefont
  {Ruan}}, \bibinfo {author} {\bibfnamefont {C.}~\bibnamefont {Liu}}, \bibinfo
  {author} {\bibfnamefont {Z.-K.}\ \bibnamefont {Guo}}, \bibinfo {author}
  {\bibfnamefont {Y.-L.}\ \bibnamefont {Wu}},\ and\ \bibinfo {author}
  {\bibfnamefont {R.-G.}\ \bibnamefont {Cai}},\ }\bibfield  {title} {\bibinfo
  {title} {{The LISA-Taiji Network: Precision Localization of Coalescing
  Massive Black Hole Binaries}},\ }\href
  {https://doi.org/10.34133/2021/6014164} {\bibfield  {journal} {\bibinfo
  {journal} {Research}\ }\textbf {\bibinfo {volume} {2021}},\ \bibinfo {pages}
  {6014164} (\bibinfo {year} {2021})},\ \Eprint
  {https://arxiv.org/abs/1909.07104} {arXiv:1909.07104 [gr-qc]} \BibitemShut
  {NoStop}%
\bibitem [{\citenamefont {Yang}\ \emph {et~al.}(2022)\citenamefont {Yang},
  \citenamefont {Han}, \citenamefont {Yun}, \citenamefont {Xu},\ and\
  \citenamefont {Luo}}]{Yang:2022cgm}%
  \BibitemOpen
  \bibfield  {author} {\bibinfo {author} {\bibfnamefont {Y.}~\bibnamefont
  {Yang}}, \bibinfo {author} {\bibfnamefont {W.-B.}\ \bibnamefont {Han}},
  \bibinfo {author} {\bibfnamefont {Q.}~\bibnamefont {Yun}}, \bibinfo {author}
  {\bibfnamefont {P.}~\bibnamefont {Xu}},\ and\ \bibinfo {author}
  {\bibfnamefont {Z.}~\bibnamefont {Luo}},\ }\bibfield  {title} {\bibinfo
  {title} {{Tracing astrophysical black hole seeds and primordial black holes
  with LISA-Taiji network}},\ }\href {https://doi.org/10.1093/mnras/stac920}
  {\bibfield  {journal} {\bibinfo  {journal} {Mon. Not. Roy. Astron. Soc.}\
  }\textbf {\bibinfo {volume} {512}},\ \bibinfo {pages} {6217} (\bibinfo {year}
  {2022})},\ \Eprint {https://arxiv.org/abs/2205.00408} {arXiv:2205.00408
  [gr-qc]} \BibitemShut {NoStop}%
\bibitem [{\citenamefont {Chen}\ \emph {et~al.}(2021)\citenamefont {Chen},
  \citenamefont {Yan}, \citenamefont {Lu}, \citenamefont {Zhao},\ and\
  \citenamefont {Ge}}]{Chen:2021sco}%
  \BibitemOpen
  \bibfield  {author} {\bibinfo {author} {\bibfnamefont {J.}~\bibnamefont
  {Chen}}, \bibinfo {author} {\bibfnamefont {C.-S.}\ \bibnamefont {Yan}},
  \bibinfo {author} {\bibfnamefont {Y.-J.}\ \bibnamefont {Lu}}, \bibinfo
  {author} {\bibfnamefont {Y.-T.}\ \bibnamefont {Zhao}},\ and\ \bibinfo
  {author} {\bibfnamefont {J.-Q.}\ \bibnamefont {Ge}},\ }\bibfield  {title}
  {\bibinfo {title} {{On detecting stellar binary black holes via the
  LISA-Taiji network}},\ }\href {https://doi.org/10.1088/1674-4527/21/11/285}
  {\bibfield  {journal} {\bibinfo  {journal} {Res. Astron. Astrophys.}\
  }\textbf {\bibinfo {volume} {21}},\ \bibinfo {pages} {285} (\bibinfo {year}
  {2021})},\ \Eprint {https://arxiv.org/abs/2201.12516} {arXiv:2201.12516
  [astro-ph.HE]} \BibitemShut {NoStop}%
\bibitem [{\citenamefont {Yang}(2021)}]{Yang:2021qge}%
  \BibitemOpen
  \bibfield  {author} {\bibinfo {author} {\bibfnamefont {T.}~\bibnamefont
  {Yang}},\ }\bibfield  {title} {\bibinfo {title} {{Gravitational-Wave Detector
  Networks: Standard Sirens on Cosmology and Modified Gravity Theory}},\ }\href
  {https://doi.org/10.1088/1475-7516/2021/05/044} {\bibfield  {journal}
  {\bibinfo  {journal} {JCAP}\ }\textbf {\bibinfo {volume} {05}},\ \bibinfo
  {pages} {044}},\ \Eprint {https://arxiv.org/abs/2103.01923} {arXiv:2103.01923
  [astro-ph.CO]} \BibitemShut {NoStop}%
\bibitem [{\citenamefont {Wang}\ \emph
  {et~al.}(2022{\natexlab{a}})\citenamefont {Wang}, \citenamefont {Jin},
  \citenamefont {Zhang},\ and\ \citenamefont {Zhang}}]{Wang:2021srv}%
  \BibitemOpen
  \bibfield  {author} {\bibinfo {author} {\bibfnamefont {L.-F.}\ \bibnamefont
  {Wang}}, \bibinfo {author} {\bibfnamefont {S.-J.}\ \bibnamefont {Jin}},
  \bibinfo {author} {\bibfnamefont {J.-F.}\ \bibnamefont {Zhang}},\ and\
  \bibinfo {author} {\bibfnamefont {X.}~\bibnamefont {Zhang}},\ }\bibfield
  {title} {\bibinfo {title} {{Forecast for cosmological parameter estimation
  with gravitational-wave standard sirens from the LISA-Taiji network}},\
  }\href {https://doi.org/10.1007/s11433-021-1736-6} {\bibfield  {journal}
  {\bibinfo  {journal} {Sci. China Phys. Mech. Astron.}\ }\textbf {\bibinfo
  {volume} {65}},\ \bibinfo {pages} {210411} (\bibinfo {year}
  {2022}{\natexlab{a}})},\ \Eprint {https://arxiv.org/abs/2101.11882}
  {arXiv:2101.11882 [gr-qc]} \BibitemShut {NoStop}%
\bibitem [{\citenamefont {Wang}\ \emph
  {et~al.}(2022{\natexlab{b}})\citenamefont {Wang}, \citenamefont {Ruan},
  \citenamefont {Yang}, \citenamefont {Guo}, \citenamefont {Cai},\ and\
  \citenamefont {Hu}}]{Wang:2020dkc}%
  \BibitemOpen
  \bibfield  {author} {\bibinfo {author} {\bibfnamefont {R.}~\bibnamefont
  {Wang}}, \bibinfo {author} {\bibfnamefont {W.-H.}\ \bibnamefont {Ruan}},
  \bibinfo {author} {\bibfnamefont {Q.}~\bibnamefont {Yang}}, \bibinfo {author}
  {\bibfnamefont {Z.-K.}\ \bibnamefont {Guo}}, \bibinfo {author} {\bibfnamefont
  {R.-G.}\ \bibnamefont {Cai}},\ and\ \bibinfo {author} {\bibfnamefont
  {B.}~\bibnamefont {Hu}},\ }\bibfield  {title} {\bibinfo {title} {{Hubble
  parameter estimation via dark sirens with the LISA-Taiji network}},\ }\href
  {https://doi.org/10.1093/nsr/nwab054} {\bibfield  {journal} {\bibinfo
  {journal} {Natl. Sci. Rev.}\ }\textbf {\bibinfo {volume} {9}},\ \bibinfo
  {pages} {nwab054} (\bibinfo {year} {2022}{\natexlab{b}})},\ \Eprint
  {https://arxiv.org/abs/2010.14732} {arXiv:2010.14732 [astro-ph.CO]}
  \BibitemShut {NoStop}%
\bibitem [{\citenamefont {Machado}\ \emph {et~al.}(2020)\citenamefont
  {Machado}, \citenamefont {Ratzinger}, \citenamefont {Schwaller},\ and\
  \citenamefont {Stefanek}}]{Machado:2019xuc}%
  \BibitemOpen
  \bibfield  {author} {\bibinfo {author} {\bibfnamefont {C.~S.}\ \bibnamefont
  {Machado}}, \bibinfo {author} {\bibfnamefont {W.}~\bibnamefont {Ratzinger}},
  \bibinfo {author} {\bibfnamefont {P.}~\bibnamefont {Schwaller}},\ and\
  \bibinfo {author} {\bibfnamefont {B.~A.}\ \bibnamefont {Stefanek}},\
  }\bibfield  {title} {\bibinfo {title} {{Gravitational wave probes of
  axionlike particles}},\ }\href {https://doi.org/10.1103/PhysRevD.102.075033}
  {\bibfield  {journal} {\bibinfo  {journal} {Phys. Rev. D}\ }\textbf {\bibinfo
  {volume} {102}},\ \bibinfo {pages} {075033} (\bibinfo {year} {2020})},\
  \Eprint {https://arxiv.org/abs/1912.01007} {arXiv:1912.01007 [hep-ph]}
  \BibitemShut {NoStop}%
\bibitem [{\citenamefont {Salehian}\ \emph {et~al.}(2021)\citenamefont
  {Salehian}, \citenamefont {Gorji}, \citenamefont {Mukohyama},\ and\
  \citenamefont {Firouzjahi}}]{Salehian:2020dsf}%
  \BibitemOpen
  \bibfield  {author} {\bibinfo {author} {\bibfnamefont {B.}~\bibnamefont
  {Salehian}}, \bibinfo {author} {\bibfnamefont {M.~A.}\ \bibnamefont {Gorji}},
  \bibinfo {author} {\bibfnamefont {S.}~\bibnamefont {Mukohyama}},\ and\
  \bibinfo {author} {\bibfnamefont {H.}~\bibnamefont {Firouzjahi}},\ }\bibfield
   {title} {\bibinfo {title} {{Analytic study of dark photon and gravitational
  wave production from axion}},\ }\href
  {https://doi.org/10.1007/JHEP05(2021)043} {\bibfield  {journal} {\bibinfo
  {journal} {JHEP}\ }\textbf {\bibinfo {volume} {05}},\ \bibinfo {pages}
  {043}},\ \Eprint {https://arxiv.org/abs/2007.08148} {arXiv:2007.08148
  [hep-ph]} \BibitemShut {NoStop}%
\bibitem [{\citenamefont {Sun}\ and\ \citenamefont
  {Zhang}(2021)}]{Sun:2020gem}%
  \BibitemOpen
  \bibfield  {author} {\bibinfo {author} {\bibfnamefont {S.}~\bibnamefont
  {Sun}}\ and\ \bibinfo {author} {\bibfnamefont {Y.-L.}\ \bibnamefont
  {Zhang}},\ }\bibfield  {title} {\bibinfo {title} {{Fast gravitational wave
  bursts from axion clumps}},\ }\href
  {https://doi.org/10.1103/PhysRevD.104.103009} {\bibfield  {journal} {\bibinfo
   {journal} {Phys. Rev. D}\ }\textbf {\bibinfo {volume} {104}},\ \bibinfo
  {pages} {103009} (\bibinfo {year} {2021})},\ \Eprint
  {https://arxiv.org/abs/2003.10527} {arXiv:2003.10527 [hep-ph]} \BibitemShut
  {NoStop}%
\bibitem [{\citenamefont {Li}\ \emph {et~al.}(2024)\citenamefont {Li},
  \citenamefont {Sun}, \citenamefont {Yan},\ and\ \citenamefont
  {Zhao}}]{Li:2023vuu}%
  \BibitemOpen
  \bibfield  {author} {\bibinfo {author} {\bibfnamefont {M.}~\bibnamefont
  {Li}}, \bibinfo {author} {\bibfnamefont {S.}~\bibnamefont {Sun}}, \bibinfo
  {author} {\bibfnamefont {Q.-S.}\ \bibnamefont {Yan}},\ and\ \bibinfo {author}
  {\bibfnamefont {Z.}~\bibnamefont {Zhao}},\ }\bibfield  {title} {\bibinfo
  {title} {{Gravitational waves from axion wave production}},\ }\href
  {https://doi.org/10.1140/epjc/s10052-024-13521-y} {\bibfield  {journal}
  {\bibinfo  {journal} {Eur. Phys. J. C}\ }\textbf {\bibinfo {volume} {84}},\
  \bibinfo {pages} {1165} (\bibinfo {year} {2024})},\ \Eprint
  {https://arxiv.org/abs/2309.08407} {arXiv:2309.08407 [hep-ph]} \BibitemShut
  {NoStop}%
\bibitem [{\citenamefont {Ding}\ \emph {et~al.}(2024)\citenamefont {Ding},
  \citenamefont {Fu}, \citenamefont {Xu},\ and\ \citenamefont
  {Zhang}}]{Ding:2024}%
  \BibitemOpen
  \bibfield  {author} {\bibinfo {author} {\bibfnamefont {K.}~\bibnamefont
  {Ding}}, \bibinfo {author} {\bibfnamefont {C.}~\bibnamefont {Fu}}, \bibinfo
  {author} {\bibfnamefont {B.}~\bibnamefont {Xu}},\ and\ \bibinfo {author}
  {\bibfnamefont {Y.-L.}\ \bibnamefont {Zhang}},\ }\bibfield  {title} {\bibinfo
  {title} {Chiral gravitational wave background in millihertz from axion-like
  fields},\ }\href {https://doi.org/10.1360/SSPMA-2024-0088} {\bibfield
  {journal} {\bibinfo  {journal} {SCIENTIA SINICA Physica, Mechanica,
  Astronomica}\ }\textbf {\bibinfo {volume} {54}},\ \bibinfo {pages} {270408}
  (\bibinfo {year} {2024})}\BibitemShut {NoStop}%
\bibitem [{\citenamefont {Babak}\ and\ \citenamefont
  {Petiteau}(2020)}]{wgLISADataChallenge2020}%
  \BibitemOpen
  \bibfield  {author} {\bibinfo {author} {\bibfnamefont {S.}~\bibnamefont
  {Babak}}\ and\ \bibinfo {author} {\bibfnamefont {A.}~\bibnamefont
  {Petiteau}},\ }\href@noop {} {\emph {\bibinfo {title} {{LISA Data Challenge
  Manual}}}},\ \bibinfo {type} {Tech. Rep.}\ \bibinfo {number}
  {LISA-LCST-SGS-MAN-002}\ (\bibinfo  {institution} {APC Paris},\ \bibinfo
  {year} {2020})\BibitemShut {NoStop}%
\bibitem [{\citenamefont {Armano}\ \emph {et~al.}(2016)\citenamefont {Armano}
  \emph {et~al.}}]{Armano:2016bkm}%
  \BibitemOpen
  \bibfield  {author} {\bibinfo {author} {\bibfnamefont {M.}~\bibnamefont
  {Armano}} \emph {et~al.},\ }\bibfield  {title} {\bibinfo {title} {{Sub-Femto-
  g Free Fall for Space-Based Gravitational Wave Observatories: LISA Pathfinder
  Results}},\ }\href {https://doi.org/10.1103/PhysRevLett.116.231101}
  {\bibfield  {journal} {\bibinfo  {journal} {Phys. Rev. Lett.}\ }\textbf
  {\bibinfo {volume} {116}},\ \bibinfo {pages} {231101} (\bibinfo {year}
  {2016})}\BibitemShut {NoStop}%
\bibitem [{\citenamefont {Babak}\ \emph {et~al.}(2021)\citenamefont {Babak},
  \citenamefont {Petiteau},\ and\ \citenamefont {Hewitson}}]{Babak:2021mhe}%
  \BibitemOpen
  \bibfield  {author} {\bibinfo {author} {\bibfnamefont {S.}~\bibnamefont
  {Babak}}, \bibinfo {author} {\bibfnamefont {A.}~\bibnamefont {Petiteau}},\
  and\ \bibinfo {author} {\bibfnamefont {M.}~\bibnamefont {Hewitson}},\
  }\bibfield  {title} {\bibinfo {title} {{LISA Sensitivity and SNR
  Calculations}},\ }\href@noop {} {\bibfield  {journal} {\bibinfo  {journal}
  {arXiv}\ } (\bibinfo {year} {2021})},\ \Eprint
  {https://arxiv.org/abs/2108.01167} {arXiv:2108.01167 [astro-ph.IM]}
  \BibitemShut {NoStop}%
\bibitem [{\citenamefont {Luo}\ \emph {et~al.}(2020)\citenamefont {Luo},
  \citenamefont {Guo}, \citenamefont {Jin}, \citenamefont {Wu},\ and\
  \citenamefont {Hu}}]{Luo:2019zal}%
  \BibitemOpen
  \bibfield  {author} {\bibinfo {author} {\bibfnamefont {Z.}~\bibnamefont
  {Luo}}, \bibinfo {author} {\bibfnamefont {Z.}~\bibnamefont {Guo}}, \bibinfo
  {author} {\bibfnamefont {G.}~\bibnamefont {Jin}}, \bibinfo {author}
  {\bibfnamefont {Y.}~\bibnamefont {Wu}},\ and\ \bibinfo {author}
  {\bibfnamefont {W.}~\bibnamefont {Hu}},\ }\bibfield  {title} {\bibinfo
  {title} {{A brief analysis to Taiji: Science and technology}},\ }\href
  {https://doi.org/10.1016/j.rinp.2019.102918} {\bibfield  {journal} {\bibinfo
  {journal} {Results Phys.}\ }\textbf {\bibinfo {volume} {16}},\ \bibinfo
  {pages} {102918} (\bibinfo {year} {2020})}\BibitemShut {NoStop}%
\bibitem [{\citenamefont {Luo}\ \emph {et~al.}(2021)\citenamefont {Luo},
  \citenamefont {Wang}, \citenamefont {Wu}, \citenamefont {Hu},\ and\
  \citenamefont {Jin}}]{Luo:2021qji}%
  \BibitemOpen
  \bibfield  {author} {\bibinfo {author} {\bibfnamefont {Z.}~\bibnamefont
  {Luo}}, \bibinfo {author} {\bibfnamefont {Y.}~\bibnamefont {Wang}}, \bibinfo
  {author} {\bibfnamefont {Y.}~\bibnamefont {Wu}}, \bibinfo {author}
  {\bibfnamefont {W.}~\bibnamefont {Hu}},\ and\ \bibinfo {author}
  {\bibfnamefont {G.}~\bibnamefont {Jin}},\ }\bibfield  {title} {\bibinfo
  {title} {{The Taiji program: A concise overview}},\ }\href
  {https://doi.org/10.1093/ptep/ptaa083} {\bibfield  {journal} {\bibinfo
  {journal} {PTEP}\ }\textbf {\bibinfo {volume} {2021}},\ \bibinfo {pages}
  {05A108} (\bibinfo {year} {2021})}\BibitemShut {NoStop}%
\bibitem [{\citenamefont {Prince}\ \emph {et~al.}(2002)\citenamefont {Prince},
  \citenamefont {Tinto}, \citenamefont {Larson},\ and\ \citenamefont
  {Armstrong}}]{Prince:2002hp}%
  \BibitemOpen
  \bibfield  {author} {\bibinfo {author} {\bibfnamefont {T.~A.}\ \bibnamefont
  {Prince}}, \bibinfo {author} {\bibfnamefont {M.}~\bibnamefont {Tinto}},
  \bibinfo {author} {\bibfnamefont {S.~L.}\ \bibnamefont {Larson}},\ and\
  \bibinfo {author} {\bibfnamefont {J.~W.}\ \bibnamefont {Armstrong}},\
  }\bibfield  {title} {\bibinfo {title} {{The LISA optimal sensitivity}},\
  }\href {https://doi.org/10.1103/PhysRevD.66.122002} {\bibfield  {journal}
  {\bibinfo  {journal} {Phys. Rev. D}\ }\textbf {\bibinfo {volume} {66}},\
  \bibinfo {pages} {122002} (\bibinfo {year} {2002})},\ \Eprint
  {https://arxiv.org/abs/gr-qc/0209039} {arXiv:gr-qc/0209039} \BibitemShut
  {NoStop}%
\bibitem [{\citenamefont {Chen}\ \emph
  {et~al.}(2024{\natexlab{b}})\citenamefont {Chen}, \citenamefont {Liu},
  \citenamefont {Zhang},\ and\ \citenamefont {Wang}}]{Chen:2024fto}%
  \BibitemOpen
  \bibfield  {author} {\bibinfo {author} {\bibfnamefont {J.}~\bibnamefont
  {Chen}}, \bibinfo {author} {\bibfnamefont {C.}~\bibnamefont {Liu}}, \bibinfo
  {author} {\bibfnamefont {Y.-L.}\ \bibnamefont {Zhang}},\ and\ \bibinfo
  {author} {\bibfnamefont {G.}~\bibnamefont {Wang}},\ }\bibfield  {title}
  {\bibinfo {title} {{Alternative LISA-TAIJI networks: Detectability of the
  Parity Violation in Stochastic Gravitational Wave Background}},\ }\href@noop
  {} {\bibfield  {journal} {\bibinfo  {journal} {arXiv preprint}\ } (\bibinfo
  {year} {2024}{\natexlab{b}})},\ \Eprint {https://arxiv.org/abs/2412.18420}
  {arXiv:2412.18420 [gr-qc]} \BibitemShut {NoStop}%
\bibitem [{\citenamefont {Cornish}\ and\ \citenamefont
  {Larson}(2001)}]{Cornish:2001qi}%
  \BibitemOpen
  \bibfield  {author} {\bibinfo {author} {\bibfnamefont {N.~J.}\ \bibnamefont
  {Cornish}}\ and\ \bibinfo {author} {\bibfnamefont {S.~L.}\ \bibnamefont
  {Larson}},\ }\bibfield  {title} {\bibinfo {title} {{Space missions to detect
  the cosmic gravitational wave background}},\ }\href
  {https://doi.org/10.1088/0264-9381/18/17/308} {\bibfield  {journal} {\bibinfo
   {journal} {Class. Quant. Grav.}\ }\textbf {\bibinfo {volume} {18}},\
  \bibinfo {pages} {3473} (\bibinfo {year} {2001})},\ \Eprint
  {https://arxiv.org/abs/gr-qc/0103075} {arXiv:gr-qc/0103075} \BibitemShut
  {NoStop}%
\bibitem [{\citenamefont {Cornish}(2002)}]{Cornish:2001bb}%
  \BibitemOpen
  \bibfield  {author} {\bibinfo {author} {\bibfnamefont {N.~J.}\ \bibnamefont
  {Cornish}},\ }\bibfield  {title} {\bibinfo {title} {{Detecting a stochastic
  gravitational wave background with the Laser Interferometer Space Antenna}},\
  }\href {https://doi.org/10.1103/PhysRevD.65.022004} {\bibfield  {journal}
  {\bibinfo  {journal} {Phys. Rev. D}\ }\textbf {\bibinfo {volume} {65}},\
  \bibinfo {pages} {022004} (\bibinfo {year} {2002})},\ \Eprint
  {https://arxiv.org/abs/gr-qc/0106058} {arXiv:gr-qc/0106058} \BibitemShut
  {NoStop}%
\bibitem [{\citenamefont {Gong}\ \emph {et~al.}(2021)\citenamefont {Gong},
  \citenamefont {Luo},\ and\ \citenamefont {Wang}}]{Gong:2021gvw}%
  \BibitemOpen
  \bibfield  {author} {\bibinfo {author} {\bibfnamefont {Y.}~\bibnamefont
  {Gong}}, \bibinfo {author} {\bibfnamefont {J.}~\bibnamefont {Luo}},\ and\
  \bibinfo {author} {\bibfnamefont {B.}~\bibnamefont {Wang}},\ }\bibfield
  {title} {\bibinfo {title} {{Concepts and status of Chinese space
  gravitational wave detection projects}},\ }\href
  {https://doi.org/10.1038/s41550-021-01480-3} {\bibfield  {journal} {\bibinfo
  {journal} {Nature Astron.}\ }\textbf {\bibinfo {volume} {5}},\ \bibinfo
  {pages} {881} (\bibinfo {year} {2021})},\ \Eprint
  {https://arxiv.org/abs/2109.07442} {arXiv:2109.07442 [astro-ph.IM]}
  \BibitemShut {NoStop}%
\bibitem [{\citenamefont {Li}\ \emph {et~al.}(2023)\citenamefont {Li} \emph
  {et~al.}}]{Li:2023szq}%
  \BibitemOpen
  \bibfield  {author} {\bibinfo {author} {\bibfnamefont {E.-K.}\ \bibnamefont
  {Li}} \emph {et~al.},\ }\bibfield  {title} {\bibinfo {title} {{GWSpace: a
  multi-mission science data simulator for space-based gravitational wave
  detection}},\ }\href@noop {} {\bibfield  {journal} {\bibinfo  {journal}
  {arXiv preprint}\ } (\bibinfo {year} {2023})},\ \Eprint
  {https://arxiv.org/abs/2309.15020} {arXiv:2309.15020 [gr-qc]} \BibitemShut
  {NoStop}%
\bibitem [{\citenamefont {Wu}\ and\ \citenamefont {Li}(2023)}]{Wu:2023bwd}%
  \BibitemOpen
  \bibfield  {author} {\bibinfo {author} {\bibfnamefont {J.}~\bibnamefont
  {Wu}}\ and\ \bibinfo {author} {\bibfnamefont {J.}~\bibnamefont {Li}},\
  }\bibfield  {title} {\bibinfo {title} {{Subtraction of the confusion
  foreground and parameter uncertainty of resolvable galactic binaries on the
  networks of space-based gravitational-wave detectors}},\ }\href
  {https://doi.org/10.1103/PhysRevD.108.124047} {\bibfield  {journal} {\bibinfo
   {journal} {Phys. Rev. D}\ }\textbf {\bibinfo {volume} {108}},\ \bibinfo
  {pages} {124047} (\bibinfo {year} {2023})},\ \Eprint
  {https://arxiv.org/abs/2307.05568} {arXiv:2307.05568 [gr-qc]} \BibitemShut
  {NoStop}%
\bibitem [{\citenamefont {Luo}\ \emph {et~al.}(2025)\citenamefont {Luo} \emph
  {et~al.}}]{Luo:2025ewp}%
  \BibitemOpen
  \bibfield  {author} {\bibinfo {author} {\bibfnamefont {J.}~\bibnamefont
  {Luo}} \emph {et~al.},\ }\bibfield  {title} {\bibinfo {title} {{Fundamental
  Physics and Cosmology with TianQin}},\ }\href@noop {} {\bibfield  {journal}
  {\bibinfo  {journal} {arXiv preprint}\ } (\bibinfo {year} {2025})},\ \Eprint
  {https://arxiv.org/abs/2502.20138} {arXiv:2502.20138 [gr-qc]} \BibitemShut
  {NoStop}%
\bibitem [{\citenamefont {Robson}\ \emph {et~al.}(2019)\citenamefont {Robson},
  \citenamefont {Cornish},\ and\ \citenamefont {Liu}}]{Robson:2018ifk}%
  \BibitemOpen
  \bibfield  {author} {\bibinfo {author} {\bibfnamefont {T.}~\bibnamefont
  {Robson}}, \bibinfo {author} {\bibfnamefont {N.~J.}\ \bibnamefont
  {Cornish}},\ and\ \bibinfo {author} {\bibfnamefont {C.}~\bibnamefont {Liu}},\
  }\bibfield  {title} {\bibinfo {title} {{The construction and use of LISA
  sensitivity curves}},\ }\href {https://doi.org/10.1088/1361-6382/ab1101}
  {\bibfield  {journal} {\bibinfo  {journal} {Class. Quant. Grav.}\ }\textbf
  {\bibinfo {volume} {36}},\ \bibinfo {pages} {105011} (\bibinfo {year}
  {2019})},\ \Eprint {https://arxiv.org/abs/1803.01944} {arXiv:1803.01944
  [astro-ph.HE]} \BibitemShut {NoStop}%
\end{thebibliography}%

\end{document}